\documentclass[11pt]{article}
\usepackage{latexsym}
\usepackage{amsfonts}

\usepackage{amssymb}
\usepackage{amsmath,theorem}
\usepackage{cite}              
\usepackage{boxedminipage}

\def\({\left(}       \def\){\right)}
\newcommand{\Id}{{1\hspace{-0.243em}\mathrm{l}}}
\let\nn=\nonumber       
        
\def\BCS{\begin{cases}}         \def\ECS{\end{cases}} 

\let\a=\alpha   \let\b=\beta    \let\g=\gamma   \let\d=\delta   
    \let\h=\eta       \let\e=\varepsilon
   \let\l=\lambda  \let\m=\mu      
      \let\x=\xi                  
     \let\o=\omega   \let\c=\chi     \let\ps=\psi    
\let\Ph=\phi    \let\PH=\Phi        \let\O=\Omega    
          \let\G=\Gamma   \let\D=\Delta 

\newcommand{\mysection}[1]{\section{#1}\setcounter{equation}{0}}

\newcommand{\bea}{\begin{eqnarray}} 
\newcommand{\eea}{\end{eqnarray}}
\newcommand{\beann}{\begin{eqnarray*}} 
\newcommand{\eeann}{\end{eqnarray*}}
\newcommand{\beq}{\begin{equation}} 
\newcommand{\eeq}{\end{equation}}
\newcommand{\ba}{\begin{array}} 
\newcommand{\ea}{\end{array}}
\newcommand{\ben}{\begin{enumerate}} 
\newcommand{\een}{\end{enumerate}}
\newcommand{\bit}{\begin{itemize}} 
\newcommand{\eit}{\end{itemize}}

\newcommand{\cD}{{\cal D}}

\newcommand{\cF}{{\cal F}}

\newcommand{\cS}{{\cal S}}

\newcommand{\cW}{{\cal W}}

\newcommand{\5}{\bar }  
\newcommand{\6}{\partial } 
\newcommand{\7}{\hat } 
\newcommand{\4}{\tilde }

\newcommand{\sfrac}[2]{\mbox{$\frac{{#1}}{{#2}}$}\,}

\newcommand{\Ii}{{\mathrm{i}}}

\newcommand{\ep}{\varepsilon}

\newcommand{\then}{\Rightarrow}

\newcommand{\uk}{\underline{k}}
\newcommand{\ok}{{\overline{k}}}
\newcommand{\ur}{\underline{m}}
\newcommand{\OR}{{\overline{m}}}
\newcommand{\UN}{{\underline{n}}}
\newcommand{\ON}{{\overline{n}}}

\newcommand{\dr}{\raise.3ex\hbox{$\stackrel{\leftarrow}{\delta}$}{}}
\newcommand{\dl}{\raise.3ex\hbox{$\stackrel{\rightarrow}{\delta}$}{}}

\newcommand{\agh}{\mathrm{af}}

\newcommand{\gh}{\mathrm{gh}}

\def\mc{\mathcal}
\def\mcd{\mathcal{D}}

\def\3{\underline}

\def\bm{\bar{\m}}
\def\ba{\bar{\a}}
\def\bL{\bar{L}}
\def\bG{\bar{G}}
\def\CW{C^{W}}

\def\stg#1 {\hspace{#1 pt}}

\def\ep{\varepsilon}
\def\gh{\mbox{gh}\,} 
\def\agh{\mbox{agh}\,}

\begin{document}

\thispagestyle{empty}

\begin{flushright}
AEI-2001-030\\
MIS-22/2001\\
TUW-01-11\\
hep-th/0104110
\end{flushright}
\vspace{1.5cm}

\begin{center}
{\Large {\bf
Superstring BRST Cohomology
}}
\end{center}

\begin{center}
{\large
Friedemann Brandt\,$^{a,b}$, Alexander Kling\,$^c$, Maximilian Kreuzer\,$^c$
}
\end{center}

{\sl 
\begin{center} 
$^a$\,Max-Planck-Institut f\"ur Mathematik in den Naturwissenschaften,\\
Inselstra\ss e 22-26, D-04103 Leipzig, Germany\\
$^b$\,Max-Planck-Institut f\"ur Gravitationsphysik
(Albert-Einstein-In\-sti\-tut),\\ 
Am M\"uhlenberg 1, D-14476 Golm, Germany\\
$^c$\,Institut f\"ur Theoretische Physik, Technische Universit\"at Wien,\\
Wiedner Hauptstra\ss e 8-10, A-1040 Vienna, Austria
\end{center}
}
\vspace{.5cm}

\begin{abstract}
We first derive all world-sheet action functionals
for NSR superstring models with (1,1) supersymmetry and
any number of abelian gauge fields, for gauge transformations
of the standard form.
Then we prove for these models that the BRST cohomology groups 
$H^g(s)$, $g<4$ (with the antifields taken into account) are
isomorphic to those of the corresponding bosonic
string models, whose cohomology is fully known.
This implies that
the nontrivial global symmetries, Noether currents,
background charges, consistent
deformations and candidate
gauge anomalies of an NSR (1,1) superstring model
are in one-to-one correspondence with their 
bosonic counterparts.
\end{abstract}

\vspace{1cm}
\begin{center}
\end{center}
\vfill

\clearpage
\setcounter{page}{1}

\mysection{Introduction and conclusion}

We present in this paper a BRST cohomological
analysis of superstring models in the NSR formulation 
\cite{Ramond:1971gb,Neveu:1971rx,Neveu:1971iv}
with local (1,1) supersymmetry \cite{Deser:1976rb,Brink:1976sc}.
The class of models under study is quite general since it
is characterized only by requirements on the field content and
the gauge symmetries. The field content is given by
the component fields of three types of supersymmetry
multiplets:
the 2d supergravity multiplet, `matter multiplets'
containing the `target space coordinates', and 
abelian gauge field multiplets.
The number of matter multiplets and gauge field multiplets
is not fixed, i.e., our results apply to any target space
dimension (1,2,\dots) and an arbitrary number (0,1,\dots) of
abelian world-sheet gauge fields.
The supersymmetry transformations are obtained from
an analysis of the Bianchi identities of
2d supergravity in presence of abelian gauge fields. 

The first part of our analysis is the determination of
all local world-sheet actions compatible with these
requirements, using the standard form of the gauge
transformations (the question whether or not these
transformations can be nontrivially deformed is a matter
covered by the investigation in the second part of the 
paper, cf.\ comments 
at the end of section \ref{result}). This 
analysis is accomplished by a cohomological
computation in the space of
local functions which do not depend on antifields
(this is possible because we use a formulation 
in which the commutator
algebra of the gauge transformations closes off-shell).
Its result has been reported and
discussed already in \cite{Brandt:2000ib}:
when abelian gauge fields are absent, the 
cohomological analysis reproduces
the general superstring action found already in \cite{Bergshoeff:1986qr};
in presence of abelian gauge fields, it yields
locally supersymmetric extensions of the purely bosonic
actions derived in
\cite{Brandt:1998pk,Brandt:1998cy} and
may be interpreted in terms of an enlarged
target space with one `frozen' extra dimension for each gauge field.
In particular there are locally supersymmetric actions
of the Born-Infeld type among these actions \cite{Brandt:2000ib}.

In the second part of the paper we analyse the 
local BRST cohomology $H(s)$ for the models 
whose world-sheet actions were determined in the
first part\footnote{The action is needed to fix 
the BRST transformations
of the antifields. $s$ denotes the BRST differential
in the jet space associated with the 
fields and antifields \cite{Barnich:2000zw}. Our analysis is
general except for a very mild assumption (invertibility)
on the ``target space metric'', see section \ref{sigmacoho}.}.
Here and throughout this paper $H(s)$ denotes
the cohomology of the BRST differential in the
space of local functions which neither depend
explicitly on the world-sheet coordinates nor
on the world-sheet differentials,
but only on the fields, antifields and their derivatives.
This cohomology is the most important one for the
models under study because
the other local BRST cohomology groups can be
easily derived from it. This is due to
the invariance of the models under world-sheet diffeomorphisms,
owing to a general property of diffeomorphism invariant theories
discussed in detail in sections 5 and 6
of \cite{Barnich:1995ap} (see also 
\cite{Brandt:1990et,Dragon:1996md,Brandt:1997mh}).

In particular, $H(s)$ yields directly the
cohomology in form-degree 2 
of $s$ modulo the ``world-sheet exterior derivative'' $d$.
\footnote{Actually $d$ is defined on the jet space of the 
fields and antifields \cite{Barnich:2000zw}.
}
This cohomology is the most relevant one for
physical applications and denoted by
$H^{g,2}(s|d)$, where $g$ specifies the ghost number sector.
Cocycles of $H^{g,2}(s|d)$ are denoted by $\omega^{g,2}$
and the cocycle condition is
\begin{equation}
s\omega^{g,2}+d\omega^{g+1,1}=0,
\label{kotzykel}
\end{equation}
where $\omega^{g+1,1}$ is some local 1-form with ghost number $g+1$.
$\omega^{g,2}$ is a coboundary in
$H^{g,2}(s|d)$ if $\omega^{g,2}=s\omega^{g-1,2}+d\omega^{g,1}$
for some local forms $\omega^{g-1,2}$ and $\omega^{g,1}$.
$H^{g,2}(s|d)$ is related to
$H(s)$ through the descent equations as explained in 
\cite{Brandt:1990et,Barnich:1995ap,Dragon:1996md,Brandt:1997mh}.
The physically interesting cohomology groups $H^{g,2}(s|d)$ are those 
with ghost numbers
$g<2$: $H^{-1,2}(s|d)$ yields the
nontrivial Noether currents and global symmetries \cite{Barnich:1995db}, 
$H^{0,2}(s|d)$ and $H^{1,2}(s|d)$ determine the
consistent deformations \cite{Barnich:1993vg}, 
background charges \cite{Brandt:1996nn}
and candidate gauge anomalies (see, e.g., \cite{Piguet:1995er}).
The corresponding cohomology groups of $s$ are
$H^g(s)$ with $g<4$. They are the objects of our second main result:
we shall prove that these cohomology groups are isomorphic to
their counterparts in the corresponding bosonic string models%
\footnote{We believe that the isomorphism
extends to all higher ghost number sectors as well
since most parts of our proof (in fact, everything except 
for the case-by-case 
study in appendix \ref{cases}) hold for all ghost numbers.}
[the bosonic model corresponding to a particular superstring model
is obtained from the latter
simply by setting all fermions to zero in the world-sheet action]. 
Furthermore, the correspondence is very explicit:
the representatives of the $s$-cohomology of a superstring model 
are simply extensions of
their ``bosonic'' counterparts, i.e., they contain the 
representatives of the $s$-cohomology
of the corresponding bosonic string model
and complete them to $s$-cocycles of the superstring model
[analogously to the superstring action itself, which contains the
bosonic string action and completes it to a locally supersymmetric one].

This result provides a 
complete characterization of
the cohomology groups $H^g(s)$, $g<4$ because the cohomology
$H(s)$ for the bosonic string models has been completely
determined in \cite{Brandt:1996gu} (ordinary bosonic
strings) and \cite{Brandt:1998cy} (bosonic strings with
world-sheet gauge fields).
Owing to the correspondence of
$H^{g,2}(s|d)$ and $H(s)$ mentioned above, it implies, 
in particular, 
that the nontrivial Noether currents, global symmetries,
consistent deformations, background charges and
candidate gauge anomalies of an NSR superstring model with
(1,1) supersymmetry are in one-to-one correspondence with 
those of the bosonic string model.
The results for the bosonic models were
derived and discussed in detail in  
\cite{Brandt:1996gu,Brandt:1996nn,Brandt:1998pk,Brandt:1998cy,Brandt:1998ri}.
We shall not repeat or summarize them here.

We find the result quite remarkable and surprising since it
means that the local (1,1)
supersymmetry of the models under study has no effect on the structure
of the cohomology at all. We note that our analysis and result applies
analogously to heterotic strings with local
(1,0) supersymmetry (by switching off one of
the supersymmetries). However, we do not expect that it
extends to superstrings with two or more local supersymmetries
of the same chirality, such
as heterotic strings with local (2,0) supersymmetry.
These supersymmetries restrict already the world-sheet
action to special backgrounds 
\cite{Alvarez-Gaume:1981hm,Gates:1984nk,Hull:1985jv}.
Accordingly, we expect that the 
local BRST cohomology
of such superstring models is ``smaller'' than the one for
corresponding bosonic strings.

The paper is organized as follows. In section~\ref{fieldcont} we specify
the field content and the gauge and BRST transformations of the
fields. In section \ref{covfields} we construct
field variables (jet space coordinates)
which are well suited for the cohomological analysis.
This involves the super-Beltrami 
parametrization for the gravitational multiplet and a construction of 
superconformal tensor fields for the matter and gauge multiplets. 
In section \ref{action} we determine
the most general action for the field content and gauge transformations
introduced before by computing $H^2(s)$ in
the space of antifield independent local functions. 
This completes the first part of our analysis.
In section \ref{antifields} we
introduce the antifields, give their
BRST transformations and extend the superconformal tensor calculus
by constructing superconformal antifield
variables. The next two sections contain the derivation of our second
result: in section \ref{sigmacoho} we define and analyse an on-shell 
BRST cohomology $H(\sigma)$; 
in section \ref{ISO} we show that $H^g(\sigma)$
is isomorphic to $H^g(s)$ and to the cohomology of the
corresponding bosonic string model when $g<4$. 
Some details of the analysis of sections \ref{sigmacoho} 
and \ref{ISO} are collected in the appendices \ref{Hs01} and 
\ref{cases}. The remaining appendices give a short summary of the derivation 
of the gauge transformations from the supergravity
Bianchi identities and contain a 
collection of the $s$-transformations of the covariant
(= superconformal) field and antifield variables.

\mysection{Field content and gauge symmetries}\label{fieldcont}

The field content of the models we are going to study is given by the 
supergravity multiplet consisting of the vielbein $e_m^{~a}$, the gravitino 
$\chi_m^\a$ and an auxiliary scalar field $S$.\footnote{$m,a,\alpha$ denote 
2$d$ world-sheet, Lorentz and spinor indices, respectively.} Furthermore we 
consider a set of scalar multipets $\{X^M,\ps_{\a}^M, F^M\}$ corresponding 
to the string ``target space coordinates'' and their superpartners and a set 
of abelian gauge multiplets $\{A^i_{m},\l_{\a}^i,\Ph^i\}$. On 
Minkowskian world-sheets all fields are real and the fermions are 
Majorana-Weyl spinors. The number of scalar multiplets 
and gauge multiplets is not specified, i.e. our approach covers any 
number of such fields. As gauge symmetries we impose world-sheet 
diffeomorphisms, local $2d$ Lorentz transformations, Weyl and 
super-Weyl transformations and of course local (1,1) world-sheet 
supersymmetry. Furthermore we require invariance under abelian 
gauge transformations of the $A^i_{m}$ and under arbitrary local shifts of 
the auxiliary field $S$. The gauge symmetries entail the corresponding ghost 
fields, which fixes the field content to 
\[
\PH^{A}=\{e_{m}^{~a},\chi_{m}^{~\a},S,X^{M},\ps_{\a}^{M},F^{M},A^i_{m},
\l_{\a}^i,\Ph^i,\x^{m},\x^{\a},C^{ab},C^W,\eta^{\a},W,c^i\},
\]
where $\x^m$ denote the world sheet diffeomorphism ghosts, $\x^\a$ are the 
supersymmetry ghosts and $C^{ab}$ is the Lorentz ghost. $C^W$ and $\eta^{\a}$ 
are the Weyl and super-Weyl ghosts, respectively. $c^i$ are the ghosts 
associated with the $U(1)$ transformations of the gauge fields and $W$ denotes 
the ghost corresponding to the local shifts of the auxiliary field $S$. The 
gauge transformation of the supergravity multiplet written as BRST 
transformations are
\begin{eqnarray}\label{sugra}
se_{m}^{~a} & = & \x^{n}\partial_{n}e_{m}^{~a}+(\partial_{m}\x^{n})e_{n}^{~a}-
                  2\Ii\x^{\a}\chi_{m}^{~\b}(\g^aC)_{\a\b}+C_{b}^{~a}e_{m}^{~b}+
                  \CW e_{m}^{~a} \nn 
\\
s\chi_{m}^{~\a} & = & \x^{n}\partial_{n}\chi_{m}^{~\a}
                     +(\partial_{m}\x^{n})\chi_{n}^{~\a}+\nabla_{m}\x^{\a}
                     -\sfrac{1}{4}\x^{\b}e_{m}^{~a}S(\g_a)_{\b}{}^{\a}
                     +\sfrac{1}{2} \CW \chi_{m}^{~\a} \nn 
\\
       &  & +\Ii\h^{\b}(\g_{m})_{\b}^{~\a}
            -\sfrac{1}{4}C^{ab}\chi_{m}^{~\b}\e_{ab}(\g_{*})_{\b}^{~\a} \nn 
\\
sS & = & \x^{n}\partial_{n}S
        -4\x^{\g}(\g_{*}C)_{\g\a}\e^{nm}\nabla_{n}\chi_{m}^{~\a}
        +\Ii\,\x^{\g}(\g^{m}C)_{\g\a}\chi_{m}^{~\a}S-\CW S+W,
\end{eqnarray}
where $C_{\a\b}$ is the charge conjugation matrix satisfying 
$-(\g^a)^T=C^{-1}(\g^a)C$. $\g_*$ is defined through 
$\g^a\g^b=\h^{ab}\Id + \e^{ab}\g_*$ and $\e^{01}=\e_{10}=1$.
$\nabla_m$ denotes the Lorentz covariant derivative
\[
\nabla_m = \partial_m - \sfrac{1}{2} \o_m^{~ab}l_{ab}
\]
in terms of the Lorentz generator $l_{ab}$ and the spin connection
\begin{eqnarray}
\o_m^{~ab} &=& E^{an}E^{bk}(\o_{[mn]k}-\o_{[nk]m}+\o_{[km]n}) \nn
\\
\o_{[mn]k} &=& e_{kd}\partial_{[n}e_{m]}^{~d}-\Ii\chi_n\g_k\chi_m, 
\qquad E_{a}^{~m}e_{m}^{~b}=\d_{a}^{~b}. 
\label{spinconnection}
\end{eqnarray}
The BRST transformations of the scalar multiplets read
\begin{eqnarray}\label{matter}
sX^{M} & = & \x^{m}\partial_{m}X^{M}+\x^{\a}\ps_{\a}^{M} \nn 
\\
s\ps_{\a}^{M} & = & \x^{m}\partial_{m}\ps_{\a}^{M}
                    -\Ii\x^{\b}(\g^mC)_{\b\a}(\partial_{m}X^{M}
                    -\chi_{m}^{~\g}\ps_{\g}^{M})+\x^{\b}C_{\b\a}F^{M} \nn 
\\
              &  &  +\sfrac{1}{4}C^{ab}\e_{ab}(\g_{*})_{\a}^{~\b}\ps_{\b}^{M}
                    -\sfrac{1}{2} C^W \ps_{\a}^{M} \nn 
\\
sF^{M} & = & \x^{m}\partial_{m}F^{M}
             +\x^{\a}(\g^{m})_{\a}^{~\b}\{\nabla_{m}\ps_{\b}^{M}
             +\Ii\chi_{m}^{~\g}(\g^nC)_{\g\b}(\partial_{n}X^{M}
             -\chi_{n}^{~\d}\ps_{\d}^{M}) \nn
\\
         & & -\chi_{m}^{~\g}C_{\g\b}F^{M}\}-C^W F^{M}.
\end{eqnarray}
The BRST transformations of the $U(1)$ multiplets are
\begin{eqnarray}\label{U1}
s\Ph^i &=& \x^{n}\partial_{n}\Ph^i
          +\x^{\a}(\g_{*})_{\a}^{~\b}\l^i_{\b}
          -C^W \Ph^i \nn
\\
s\l^i_{\b}&=& \x^{n}\6_{n}\l^i_{\b}
             + \x^{\a}{\big (}\Ii(\g_{*}C)_{\a\b}\e^{mn}
               (\6_m A^i_n
                +\c_m\g_n\l^i-\Ii\c_n\g_*C\c_m\phi^i) \nn
\\
       &  &  - \Ii(\g_{*}\g^{m}C)_{\a\b}
               (\6_m\Ph^i-\c_m \g_* \l^i ) 
             + \Ii(\g_{*}C)_{\a\b}S\Ph^i {\big )} \nn
\\
       &  & + \sfrac{1}{4} C^{ab}\e_{ab}(\g_{*})_{\b}^{~\g}\l^i_{\g} 
            + 2\h^{\a}(\g_{*}C)_{\a\b}\Ph^i 
            - \sfrac{3}{2} C^W \l^i_{\b} \nn
\\
sA^i_{m} &=& \x^{n}\partial_{n}A^i_{m}+(\partial_{m}\x^{n})A^i_{n}
            +\partial_{m}c^i \nn
\\
       & &  -2\Ii\,\x^{\a}\c_{m}^{~\b}(\g_* C)_{\b\a}\Ph^i
            -\x^{\a}(\g_m)_{\a}^{~\b}\l^i_{\b}.
\end{eqnarray}
These transformations were obtained by analyzing the $2d$ supergravity 
algebra in presence of the scalar matter and gauge multiplets~\cite{K} 
analogously to the superspace analysis of~\cite{Howe:1979ia}. A short 
summary of the analysis is given in appendix \ref{gaugealgebra}. In the 
supergravity sector we used the constraints 
\begin{equation}\label{gravconstr}
T_{\a\b}{}^a = 2\Ii(\g^aC)_{\a\b},\quad T_{ab}{}^c=T_{\a\b}{}^\g=0
\end{equation}
and in the $U(1)$ sector
\begin{equation}\label{u1constr}
F^i_{\a\b}=2\Ii(\g_*C)_{\a\b}\Ph^i.
\end{equation}
All constraints are conventional, i.e., can be achieved by redefinitions of 
the connections. The transformations of the ghosts are such that the BRST 
differential $s$ squares to zero, 
\begin{eqnarray}\label{ghosts}
s\x^{n} & = & \x^{m}\partial_{m}\x^{n}+\Ii\x^{\a}\x^{\b}(\g^nC)_{\a\b} \nn 
\\
s\x^{\a} & = & \x^{n}\partial_{n}\x^{\a}
              -\Ii\x^{\g}\x^{\b}(\g^mC)_{\b\g}\chi_{m}^{~\a}
              -\sfrac{1}{4}C^{ab}\x^{\b}\e_{ab}(\g_{*})_{\b}^{~\a}
              +\sfrac{1}{2} C^W \x^{\a} \nn 
\\
sC^{ab} & = & \x^{m}\partial_{m}C^{ab}
              -\sfrac{\Ii}{4}\x^{\a}\x^{\b}S(\g_*C)_{\a\b}\ep^{ab}
              -\Ii\x^{\a}\x^{\b}(\g^mC)_{\b\a}\omega_{m}^{~ab}
              -2\h^{\b}\x^{\a}(\g_{*}C)_{\a\b}\e^{ab} \nn 
\\
sC^W & = & \x^{n}\partial_{n}C^W+2\h^{\b}\x_{\b} \nn 
\\
s\h^{\a} & = & \x^{n}\partial_{n}\h^{\a}
              -\sfrac{1}{4} C^{ab}\h^{\b}\e_{ab}(\g_{*})_{\b}^{~\a}
              +\Ii\x^{\b}(\g^{n})_{\b}^{~\a}\(\sfrac{1}{2}\partial_{n}C^W
              -\h^{\g}(\chi_{n}C)_{\g}\) \nn
\\
   & & 
              -\sfrac{1}{2} C^W \h^{\a}+\x^{\a}W \nn 
\\   
sW & = & \x^{n}\partial_{n}W
        -4\Ii\x^{\b}(\g^{m}C)_{\b\a}\(\nabla_{m}\h^{\a}
        -\sfrac{1}{4}\chi_{m}^{~\a}W
        -\sfrac{i}{2}\chi_{m}^{~\g}(\g^{n})_{\g}^{~\a}(\partial_{n}C^W)\)\nn
\\  
   & &  -4\x^{\b}\chi_{m}^{~\a}(\g^{m} \g^{n} C)_{\a \b}\h^{\g}(\chi_{n}C)_{\g}
        -C^W W \nn
\\
sc^i &=& \x^m\partial_mc^i+\Ii\x^\a\x^\b(\g_*C)_{\a\b}\phi^i
         -\Ii\x^\a\x^\b(\g^mC)_{\a\b}A_m^i.
\end{eqnarray}
We remark that the use of Weyl, super-Weyl and Lorentz
transformations, as well as the
shift symmetry associated with the auxiliary field $S$
are artefacts of the formulation and disappear in an
equivalent formulation based on a
Beltrami parametrization of the world-sheet zweibein
(see sections \ref{covfields} and \ref{action}).
Of course we could have used the Beltrami approach from the
very beginning, but we decided to start from
the more familiar formulation presented above. 

\mysection{Superconformal tensor calculus}\label{covfields}
 
The first part of our cohomological analysis consists in
the construction of a suitable ``basis''
for the fields and their derivatives
(more precisely: suitable
coordinates of the jet space
associated with the fields).
The goal is to find a basis 
$\{u^\ell,v^\ell,w^I\}$ with as many $s$-doublets
$(u^\ell,v^\ell)$ as possible and complementary (local) variables
$w^I$ such that $sw^I$ can be expressed
solely in terms of the $w$'s, i.e.,
\begin{equation}
s u^\ell = v^\ell, \quad s w^I = r^I(w).
\label{uvw}
\end{equation}
On general grounds, such a basis is related to a
tensor calculus
\cite{Brandt:1997mh,Brandt:1999iu,Brandt:2001tg}.
In the present case the tensor calculus is a
superconformal one, generalizing the
conformal tensor calculus in bosonic string models found in
\cite{Brandt:1996gu} (see also \cite{Brandt:1998cy}).
The $w$'s with ghost number 1 are specific ghost variables
corresponding to the superconformal algebra, the $w$'s with
ghost number 0 are ``superconformal tensor fields'' on which this algebra
is represented.

\subsection{Super-Beltrami parametrization}

The superconformal structure of the models under 
consideration is related to the supersymmetric generalization of the 
so-called Beltrami parametrization~\cite{Delduc:1990gn,Grimm:1990ju}. 
Beltrami differentials parametrize conformal classes of $2d$ metrics, 
and this makes them natural quantities to be used as basic variables 
in the present context.
Since Beltrami differentials change 
only under world-sheet reparametrizations but not under
Weyl or Lorentz transformations, their use leads to a simpler formulation
of the models under study (cf.\ remarks at the
end of section \ref{fieldcont}, and in section \ref{action}).
In the following we choose a Euclidean notation and parametrize 
the worldsheet with independent variables $z$ and $\bar z$ rather than 
with light cone coordinates, because this simplifies the notation and 
avoids some factors of $\Ii$.\footnote{Note that reality conditions of 
spinors are subtle after Wick rotation to Euclidean space: 
In our left-right symmetric case of (1,1) supersymmetry we could 
define $(\ps)^*=\bar\ps$ and work with manifestly real actions, 
but obviously this would not be possible for heterotic theories. 
This is, however, irrelevant in our algebraic context.}

As it is not hard to guess the supersymmetric generalization of the Beltrami 
parametrization involves in addition to the bosonic Beltrami differential $\m$ 
a fermionic partner $\a$, the Beltramino. 
The starting point is the parametrization of the vielbein
\begin{eqnarray}
e^{z}&=&(dz+d\bar z\m_{\bar z}^{~z})e_{z}^{~z} \nn\\
e^{\bar z}&=&(d\bar z+dz\m_{z}^{~\bar z})e_{\bar z}^{~\bar z}.
\end{eqnarray}
The coefficients $\m_{\bar z}^{~z}$ and $\m_{z}^{~\bar z}$ are the 
Beltrami differentials 
\begin{eqnarray}
\m &:=& \m_{\bar z}^{~z}=\frac{e_{\bar z}^{~z}}{e_{z}^{~z}},\nn \\
\bm &:=& \m_{z}^{~\bar z}=\frac{e_{z}^{~\bar z}}{e_{\bar z}^{~\bar z}},
\end{eqnarray}
whereas the factors $e_{z}^{~z}$ and $e_{\bar z}^{~\bar z}$ are referred to 
as conformal factors. One should note that the Beltrami differentials 
transform under diffeomorphisms but do not change under Weyl or Lorentz 
transformations. The latter
``structure group transformations'' are carried solely by 
the conformal factors
which form $s$-doublets $(u^\ell,v^\ell)$ with ghost variables substituting
(in the new basis) for the Lorentz ghost and the Weyl ghost.

The fermionic superpartners of the Beltrami differentials are suitable 
combinations of the gravitino fields
\begin{eqnarray}
\a &:=& \sqrt{\sfrac{8}{e_{z}^{~z}}}
        \(\chi_{\bar z}^{~2}-\m\chi_{z}^{~2}\) \nn
\\
\ba &:=& \sqrt{\sfrac{8}{e_{\bar z}^{~\bar z}}}
         \(\chi_{z}^{~1}-\bm\chi_{\bar z}^{~1}\).
\end{eqnarray}
The Beltraminos are also invariant under structure group transformations. 
Especially they do not change under super-Weyl transformations. Again one can 
find complementary combinations of the gravitinos 
forming $s$-doublets with ghost variables that substitute for the 
super-Weyl ghosts. The fact that Weyl, Lorentz and super-Weyl
ghosts (and not just their derivatives) occur
in $s$-doublets as we just described reflects that 
Weyl, Lorentz and super-Weyl invariance are artefacts 
of the formulation.

The Beltrami parametrization involves also a redefinition of the 
diffeomorphism ghosts, sometimes called the Beltrami ghost fields. This again 
has to be supplemented with a redefinition of the supersymmetry ghosts. The 
new ghost variables, which replace the diffeomorphism ghosts $\x^{z}$ and 
$\x^{\bar z}$ and the supersymmetry ghosts $\x^{1}$ and $\x^{2}$ are 
\begin{eqnarray}
\h &:=& (\x^{z}+\m\x^{\bar z}) \nn\\
\bar\h &:=& (\x^{\bar z}+\bm\x^{z}) \nn\\
\e &:=& \sfrac 12(\hat \x^{2}+\x^{\bar z}\a),~~~~
\hat \x^{2}:=\sqrt{\sfrac{8}{e_{z}^{~z}}}\x^{2} \nn\\ 
\bar\e &:=& \sfrac 12(\hat \x^{1}+\x^{z}\ba),~~~~
\hat \x^{1}:=\sqrt{\sfrac{8}{e_{\bar z}^{~\bar z}}}\x^{1}
\label{redefinitions} 
\end{eqnarray}
In terms of the new ghost variables the BRST transformations 
of ``right-moving'' and ``left-moving'' quantities
decouple from each other~\cite{Delduc:1990gn},
\begin{eqnarray}
s\m &=& \(\bar\partial-\m\partial+(\partial\m)\)\h+\a\e \nn
\\
s\a &=&  \(2\bar\partial-2\m\partial+(\partial\m)\)\e+\h\partial\a+
         \sfrac{1}{2}\a\partial\h \nn
\\
s\h &=& \h\partial\h - \e\e \nn
\\
s\e &=& \h\partial\e-\sfrac{1}{2}\e\partial\h,
\label{sbt}\end{eqnarray}
with analogous transformations for the right movers. 

\subsection{Superconformal ghost variables and algebra}\label{confstructure}

We have now paved the road for the construction of
field variables $\{u^\ell,v^\ell,w^I\}$ fulfilling (\ref{uvw}).
In fact we have already identified some $s$-doublets
$(u^\ell,v^\ell)$, namely the $u$'s given by the conformal factors 
and their fermionic counterparts and the corresponding $v$'s given 
by ghost fields substituting in the new basis for
the Weyl, Lorentz and super-Weyl ghosts.
Furthermore, the field $S$ obviously forms an $s$-doublet with
a ghost field substituting for $W$. The derivatives of
these $u$'s and $v$'s form $s$-doublets as well.
The Beltrami differentials $\mu,\bar\mu$ and their derivatives are
$u$'s too. From (\ref{sbt}) one observes that $s\mu$ and 
$s\bar\mu$ contain derivatives $\bar\partial \eta$
and $\partial {\bar\eta}$ and 
of the reparametrization ghosts, respectively. 
Taking derivatives of these transformations, one sees that
the $m$-th derivatives of the Beltrami differentials pair off with 
ghost variables that substitute in the new basis for all 
$(m+1)$-th derivatives of the reparametrization ghosts except for 
$\partial^{m+1} \eta$ and $\bar\partial^{m+1} {\bar\eta}$. 
Analogously, the $s$-transformations of the Beltraminos contain 
derivatives $\bar\partial\ep$ and $\partial\bar\ep$ of the supersymmetry 
ghosts. Thus the $m$-th derivatives of $\alpha$
and $\bar\alpha$ pair off with 
ghost variables substituting for all $(m+1)$-th
derivatives of $\ep$ and $\bar\ep$
except for $\partial^{m+1}\ep$ and $\bar\partial^{m+1}\bar\ep$. 
We introduce the following notation for those ghost variables
which do not sit in $s$-doublets:
\begin{equation}\label{genconn}
\{C^{N}\}=\{\h^{p},\bar{\h}^{p},\e^{p+\sfrac{1}{2}},
\bar{\e}^{p+\sfrac{1}{2}} : p=-1,0,1,\dots \},
\end{equation}
with
\begin{eqnarray}\label{newgh2}
\h^{p} &=& \frac{1}{(p+1)!}\,\partial^{p+1}\h \nn
\\
\bar{\h}^{p} &=& \frac{1}{(p+1)!}\,\bar\partial^{p+1}\bar{\h} \nn
\\
\e^{p+\sfrac{1}{2}} &=& \frac{1}{(p+1)!}\,\partial^{p+1}\e \nn
\\
\bar{\e}^{p+\sfrac{1}{2}} &=& \frac{1}{(p+1)!}\,\bar\partial^{p+1}\bar{\e}.
\end{eqnarray}
These ghost variables fulfill the requirement imposed
in (\ref{uvw}) on $w$'s. Indeed,
using (\ref{sbt}), one easily computes their $s$-transformations:
\begin{eqnarray}
s\h^{p}&=&-\sfrac{1}{2}\h^{q}\h^{r}f_{rq}^{~~p}+
\sfrac{1}{2}\e^{a}\e^{b}f_{ab}^{~~p} 
\nn \\
 &=& \sfrac{1}{2}\h^{q}\h^{r}(r-q)\d_{r+q}^{p}
     -\sfrac{1}{2}\e^{a}\e^{b}2\d_{a+b}^{p} \\
s\e^{a}&=&-\sfrac{1}{2}\h^{p}\e^{c}f_{cp}^{~~a}
+\sfrac{1}{2}\e^{c}\h^{p}f_{pc}^{~~a} 
\nn \\
 &=&  -\e^{c}\h^{p}\(\frac{p}{2}-c\)\d_{p+c}^{a}.
\end{eqnarray}
The $f$'s which occur in these transformations are
the structure constants of a graded commutator algebra
of operators $\D_{N}$ to be represented on
tensor fields constructed of the component fields of the matter
and $U(1)$ multiplets,
\begin{equation}
\{\D_{N}\}=\{L_{p},\bL_{p},G_{p+\sfrac{1}{2}},
             \bG_{p+\sfrac{1}{2}}:p=-1,0,1,\dots \}.   
\label{operators}\end{equation}
This graded commutator algebra is nothing 
but the NS superconformal algebra
\begin{equation}\label{suvirasoro}
[L_{p},L_{q}]=(p-q)L_{p+q},~~~~\{G_{a},G_{b}\}=2L_{a+b},~~~~ 
[L_{p},G_{a}]=\(\frac{p}{2}-a\)G_{p+a},
\end{equation}
with the analogous formulas for the $\bL$'s and $\bG$'s and the usual property 
that the holomorphic and antiholomorphic generators (anti-)commute,
\begin{eqnarray*}
[L_{p},\bar L_{q}]=0,&\quad &\{ G_{a},\bG_{b} \}=0,
\\~
[L_{p},\bar G_{a}]=0,&\quad &[\bL_{p},G_{a}]=0.
\end{eqnarray*}
The representation of this algebra on superconformal
tensor fields, and the explicit construction of these tensor fields,
will be given in the following subsection.

\subsection{Superconformal tensor fields}\label{covmatter}

We shall now summarize the representation of the
algebra (\ref{suvirasoro}) on superconformal tensor fields constructed
of the fields and their derivatives (the representation
on antifields is discussed in section \ref{antifields}) such
that the BRST transformation of these tensor fields
reads\footnote{$\mc{T}$ stands for any of these 
superconformal tensor fields;
$\h$'s and $\e$'s are the
ghost variables (\ref{newgh2}).}
\begin{equation}
s \mc{T} = \sum_{p\ge-1}\(\h^{p}L_{p}+\bar{\h}^{p}\bL_{p}+
\e^{p+\sfrac{1}{2}}G_{p+\sfrac{1}{2}}+
\bar{\e}^{p+\sfrac{1}{2}}\bG_{p+\sfrac{1}{2}}\)\mc{T}.  
\label{tensortrafo}\end{equation}
The superconformal
tensor fields corresponding to the fields $X^M$, $\ps^M_\alpha$, 
$F^M$ and their derivatives are denoted by $X^M_{m,n}$,
$\ps^M_{m,n}$, $\bar\ps^M_{m,n}$, $F^M_{m,n}$ ($m,n\in\{
0,1,2,\dots\}$). Here
the subscripts $m,n$ denote the number of operations
$L_{-1}$ and $\bL_{-1}$ acting on $X^M_{0,0}$,
$\ps^M_{0,0}$, $\bar\ps^M_{0,0}$, $F^M_{0,0}$, respectively
($L_{-1}$ and $\bL_{-1}$ will be identified with covariant
derivatives, see below),
\beann
&
X^M_{0,0}\equiv X^M,\ 
\ps^{M}_{0,0}\equiv 
(e_{z}^{~z}/2)^{\sfrac{1}{2}}\ps_{2}^{M},\
\bar\ps^{M}_{0,0}\equiv (e_{\bar z}^{~\bar z}/2)^{\sfrac{1}{2}}
\ps_{1}^{M},\
F^M_{0,0}\equiv \sfrac{1}{2}(e_{z}^{~z})^{\sfrac{1}{2}}
(e_{\bar z}^{~\bar z})^{\sfrac{1}{2}}F^{M},
&
\\
&
X^M_{m,n}=
(L_{-1})^{m}(\bL_{-1})^{n}X^M_{0,0}\quad
(m,n\in\{0,1,2,\dots\})\quad \mbox{etc.}
&
\eeann
The representation
on these tensor fields can be inductively deduced from the
algebra (\ref{suvirasoro}) using that
all operations $L_{m}$, $\bL_{m}$, $G_{a}$, $\bG_{a}$ vanish
on $X^M_{0,0}$ except for 
$L_{-1}$, $\bL_{-1}$, $G_{-1/2}$ and $\bG_{-1/2}$,
with $G_{-1/2}X^M_{0,0}=\ps^M_{0,0}$ and 
$\bG_{-1/2}X^M_{0,0}=\bar\ps^M_{0,0}$
(as can be read off from $sX^M$). This gives on $X^M_{m,n}$:
\begin{eqnarray}
L_{p}X^{M}_{m,n}  &  =  &  \begin{cases}
\frac{m!}{(m-p-1)!}X^{M}_{m-p,n}  &\textrm{ for $p<m$} \\
0  & \textrm{ for $p\ge m$} \end{cases} \nn 
\\
\bL_{q}X^{M}_{m,n}  &  =  & \BCS
\frac{n!}{(n-q-1)!}X^{M}_{m,n-q}  & \textrm{for $q<n$} \\
0  & \textrm{ for $q\ge n$} \ECS \nn 
\\
G_{p+\sfrac{1}{2}}X^{M}_{m,n}  &  =  &  \BCS
\frac{m!}{(m-p-1)!}\ps^{M}_{m-p-1,n} & \textrm{ for $p<m$} \\
0 & \textrm{for $p\ge m$} \ECS \nn
\\
\bG_{q+\sfrac{1}{2}}X^{M}_{m,n}  &  =  &  \BCS
\frac{n!}{(n-q-1)!}\bar \ps^{M}_{m,n-q-1} &\textrm{ for $q<n$} \\
0 & \textrm{ for $q\ge n$} \ECS \nn
\end{eqnarray}
The action on the other fields is then easily obtained using
\[
[L_{p},G_{-\sfrac{1}{2}}]=\sfrac{1}{2}\(p+1\)G_{p-\sfrac{1}{2}},~~~~
\{G_{p+\sfrac{1}{2}},G_{-\sfrac{1}{2}}\}=2L_{p}
\]
and the analogous formulas for $\bL$ and $\bG$ in (\ref{suvirasoro}).
One obtains
\begin{eqnarray}
L_{p}\ps^{M}_{m,n} & = & \begin{cases}\frac{m!}{(m-p)!}
           \(m-p+\sfrac{1}{2}(p+1)\)\ps^{M}_{m-p,n} & \textrm{ for $p\le m$}\\
           0 & \textrm{ for $p> m$}
\end{cases} \nn
\\
G_{p+\sfrac{1}{2}}\ps^{M}_{m,n} & = & \BCS
\frac{m!}{(m-p-1)!}X^{M}_{m-p,n} &\textrm{ for $p<m$} \\
0 & \textrm{ for $p\ge m$} \ECS \nn
\\
\bG_{q+\sfrac{1}{2}} \ps_{m,n}^{M} &=& \BCS
-\frac{n!}{(n-q-1)!} F^{M}_{m,n-q-1} & \textrm{ for $q<n$} \\
0 & \textrm{for $q\ge n$} \ECS \nn\\
\bL_{q}\ps^{M}_{m,n} & = & \BCS
\frac{n!}{(n-q-1)!}\ps^{M}_{m,n-q}  & \textrm{for $q<n$} \\
0  & \textrm{ for $q\ge n$} \ECS \nn 
\\
L_{p}F^{M}_{m,n} & = & \BCS
\frac{m!}{(m-p)!}\(m-p+\sfrac{1}{2}(p+1)\)F^M_{m-p,n}&\textrm{ for $p\le m$}\\
0 & \textrm{for $p> m$} \ECS \nn
\\
G_{p+\sfrac{1}{2}} F^{M}_{m,n} & = & \BCS
\frac{m!}{(m-p-1)!} \, \bar \ps^{M}_{m-p,n} &\textrm{ for $p<m$} \\
0 & \textrm{for $p\ge m$} \ECS \nn
\end{eqnarray}
and analogous formulas 
for $L$'s, $G$'s, $\bL$'s and $\bG$'s acting on $\bar \ps^{M}_{m,n}$,
and $\bL$'s and $\bG$'s acting on $F^{M}_{m,n}$.

The relation to the fields and their derivatives is established
by identifying the operations $L_{-1}$ and $\bL_{-1}$ with
covariant derivatives $\cD$ and $\bar\cD$ along
the lines of \cite{Brandt:1997mh},
\begin{eqnarray}
L_{-1}\equiv
\cD=\frac{1}{1-\mu\bar\mu}\Big[
\6-\bar\mu\bar\6-\sum_{p\geq 0}(\bar M^p\bar L_p-\bar\mu M^p L_p)
-\sum_{a}(\bar A^{a}\bar G_{a}
-\bar\mu A^{a} G_{a})
\Big]
\nn
\\
\bar L_{-1}\equiv
\bar\cD=\frac{1}{1-\mu\bar\mu}\Big[
\bar\6-\mu\6-\sum_{p\geq 0}(M^pL_p-\mu \bar M^p \bar L_p)
-\sum_{a}(A^{a}G_{a}
-\mu \bar A^{a} \bar G_{a})
\Big]
\label{covdevs}
\end{eqnarray}
where
\begin{eqnarray*}
&
M^p=\sfrac{1}{(p+1)!}\6^{p+1}\mu,\quad
\bar M^p=\sfrac{1}{(p+1)!}\bar\6^{p+1}\bar\mu,
&
\\
&
A^{p+\sfrac 12}=\sfrac{1}{(p+1)!2}\6^{p+1}\alpha,\quad
\bar A^{p+\sfrac 12}=\sfrac{1}{(p+1)!2}\bar\6^{p+1}\bar\alpha.
&
\end{eqnarray*}
One readily checks that these formulas result in local
expressions for the superconformal tensor fields and their
$s$-transformations. Introducing the following notation for the 
lowest weight superconformal matter fields
\begin{equation}
X^M\equiv X^M_{0,0}\, ,\quad
\ps^M\equiv \ps^M_{0,0}\, ,\quad
\bar\ps^M\equiv \bar\ps^M_{0,0}\, ,\quad
\7F^M\equiv F^M_{0,0}\, ,  
\label{notation1}
\end{equation}
one gets in particular the following supercovariant derivatives
\begin{eqnarray}
\cD X^M&=&\frac{1}{1-\mu\bar\mu}\Big[
(\6-\bar\mu\bar\6)X^M-\sfrac 12\bar\alpha\bar\ps^M
+\sfrac 12\bar\mu\alpha\ps^M\Big]
\nn\\
\cD \ps^M&=&\frac{1}{1-\mu\bar\mu}\Big[
(\6-\bar\mu\bar\6)\ps^M
+\sfrac 12\bar\mu(\6\mu)\ps^M
+\sfrac 12\bar\alpha \hat F^M+\sfrac 12\bar\mu\alpha\cD X^M\Big]
\nn\\
\bar\cD \ps^M&=&\frac{1}{1-\mu\bar\mu}\Big[
(\bar\6-\mu\6)\ps^M
-\sfrac 12(\6\mu)\ps^M-\sfrac 12\alpha\cD X^M
-\sfrac 12\mu\bar\alpha \hat F^M
\Big]
\label{lowcovdevs}
\end{eqnarray}
and analogous expressions for $\bar\cD X^M$,
$\bar\cD \bar\ps^M$ and $\cD \bar\ps^M$.
We do not spell out higher order covariant
derivatives explicitly because it turns out that
they do not contribute nontrivially to the cohomology.
The BRST transformations of the superconformal
tensor fields are summarized
in appendix \ref{transformations1}.

The construction of the superconformal tensor fields
arising from the gauge multiplets is similar, once one
has identified the suitable ghost variables and
the lowest order tensor fields. The
gauge fields $A_{m}^i$ and their
symmetrized derivatives $\6_{(m_1}\dots\6_{m_k}A_{m_{k+1})}^i$
($k=1,2,\dots$) form $s$-doublets with ghost variables
that substitute for all the derivatives of the ghosts $c^i$.
Therefore one expects that only the undifferentiated
ghosts $c^i$ give rise to $w$-variables. 
Promising candidates for these $w$-variables are
ghost variables $C^i$ of the same form
as in the purely bosonic case \cite{Brandt:1998cy},
\begin{eqnarray}
C^i=c^i+\x^{m}A^i_{m}\ .
\end{eqnarray}
The $s$-transformations of the gauge fields, written in terms
of $C^i$, and of the $C^i$ themselves read
\begin{eqnarray}
s A^i_{m} &=& \x^{n}(\partial_{n}A^i_{m}-\partial_{m}A^i_{n})
                +\partial_{m}C^i-\x^{\a}\chi_{m}^{~\b}F^i_{\a\b}
                -\x^{\a}e_{m}^{~a}F^i_{a\a} \nn
\\
s C^i &=& \x^{m}\x^{n}(\partial_{m}A^i_{n}
                  -\partial_{n}A^i_{m})
                  +\sfrac{1}{2}\x^{\a}\x^{\b}F^i_{\a\b}
                  +\x^{m}\x^{\a}\chi_{m}^{~\b}F^i_{\a\b}
                  +\x^{m}\x^{\a}F^i_{m\a}
\label{sC}
\end{eqnarray}
where we used notation of appendix \ref{gaugealgebra}.
Since we expect $C^i$ to count among the $w$'s,
its $s$-transformation should involve only $w$'s again, see (\ref{uvw}).
This suggests a strategy to determine the
superconformal tensor fields corresponding to the 
undifferentiated fields $\Ph^i$, $\l^{i}_\alpha$ and
to the field strengths of $A^i_m$:
one tries to rewrite $sC^i$ in (\ref{sC}) in terms of the
ghost variables (\ref{newgh2}) and to
read off from the result the sought superconformal tensor
fields. This strategy turns out to be successful; one obtains
\[
s C^i = \h\bar{\h}F^i_{0,0}+\h\bar{\e}\l^i_{0,0}
                 +\bar{\h}\e{\bar\l}{}^i_{0,0}+\e\bar{\e}\Ph^i_{0,0}
\]
where  
\begin{eqnarray}
\Ph^{i}_{0,0} &=& \sqrt{e_{z}^{~z}e_{\bar z}^{~\bar z}}\Ph^i \nn
\\
\l^{i}_{0,0} &=& \sqrt{\sfrac{e_{\bar z}^{~\bar z}}{2}}
               \(- e_{z}^{~z}\l^i_{2}+\chi_{z}^{2}\Ph^i \) \nn
\\
{\bar\l}{}^{i}_{0,0}&=& \sqrt{\sfrac{e_{z}^{~z}}{2}}
                \(e_{\bar z}^{~\bar z}\l^i_{1}+\chi_{\bar z}^{1}\Ph^i\) \nn
\\
F^{i}_{0,0}&=&\frac{1}{1-\m\bar\m}\(\sfrac{1}{2}\e^{mn}(\partial_{m}A^i_{n}
             -\partial_{n}A^i_{m})+\sfrac{1}{2}\m\bar\a\l^i
             -\sfrac{1}{2}\bar\m\a\bar\l^i-\sfrac{1}{4}\a\bar\a\Ph^i\). 
\end{eqnarray}
An explicit computation shows that the $s$-transformations
of these quantities are indeed of the desired form (\ref{tensortrafo}),
with
\begin{equation}
\l^{i}_{0,0}=G_{-\sfrac{1}{2}}\Ph^{i}_{0,0},\qquad 
{\bar\l}{}^{i}=\bG_{-\sfrac{1}{2}}\Ph^{i}_{0,0}, \qquad
F^{i}_{0,0}=\bG_{-\sfrac{1}{2}}G_{-\sfrac{1}{2}}\Ph^{i}_{0,0}.
\end{equation}
It is now straightforward to construct, along the previous lines, variables
$\Ph^{i}_{m,n}$, $\l^{i}_{m,n}$, ${\bar\l}{}^{i}_{m,n}$,
$F^{i}_{m,n}$ on which
the algebra (\ref{suvirasoro}) is represented
and (\ref{tensortrafo}) and (\ref{covdevs}) hold.
We do not spell out these tensor fields (with $m$
or $n$ different from 0) explicitly because it turns out that
they do not contribute nontrivially to the cohomology.
The resulting BRST transformations are summarized
in appendix \ref{transformations1} too.

We introduce the following notation for the lowest order (i.e.\ lowest 
weight, see below) superconformal tensor fields arising from the gauge 
multiplet:
\begin{equation}
\hat\Ph^i\equiv \Ph^{i}_{0,0}\, ,\quad
\lambda^i\equiv \l^{i}_{0,0}\, , \quad
\bar\lambda^i\equiv \bar\l^{i}_{0,0}\, , \quad
\cF^i\equiv F^{i}_{0,0}\, .
\label{notation}
\end{equation}
Again tensor fields of higher order will be denoted
by $\cD \hat\Ph^i$, $\bar\cD \hat\Ph^i$, $\cD\bar\cD \hat\Ph^i$ etc. but 
as already stated above their explicit form will not be needed.

\mysection{Action}\label{action}

We shall now determine the most general action for the field content 
and gauge transformations specified in 
section \ref{fieldcont}. The action has vanishing ghost number and is 
independent of antifields. Furthermore the requirement that the action be 
gauge invariant translates into BRST invariance up to surface terms.
The integrands of the world-sheet actions we are looking for are thus the 
antifield independent solutions $\omega^{0,2}$ of equation (\ref{kotzykel}).
They are related through the descent equations to the solutions of
\begin{eqnarray}
&s\omega=0,\quad \omega\neq s\hat\omega,& \nn\\
&\gh(\omega)=2,\quad \agh(\omega)=\agh(\hat\omega)=0&
\label{cc}
\end{eqnarray}
where $\gh$ is the ghost number and $\agh$ is the antifield number 
(=``antighost number'', see section~\ref{antifields} for the definition). 
In the previous section we have constructed a basis for the fields and 
their derivatives satisfying the requirements of (\ref{uvw}). 
By standard arguments this implies that $\omega$ and $\hat\omega$ can be 
assumed to depend only on the $w^I$, i.e., on superconformal tensor 
and ghost fields introduced in section~\ref{covfields}.%
\footnote{The $u$'s and $v$'s contribute
only ``topologically'' via the de Rham cohomology of the zweibein manifold
to the $s$-cohomology, cf.\ theorem 5.1 of \cite{Barnich:1995ap}.
In particular they do not contribute nontrivially to the solutions
of (\ref{cc}).}
Furthermore we can restrict the investigation
to functions $\omega$ and $\hat\omega$ with vanishing
``conformal weights''
by an argument used already in \cite{Brandt:1996gu,Brandt:1998cy}:
we extend the 
definition of $L_0$ and $\bar L_0$ to all $w$'s (including
the ghost variables) by
\begin{equation}
\Big\{s\, ,\, \frac{\6}{\6(\6\eta)}\,\Big\}w^I
=L_0\,w^I\ ,\ 
\Big\{s\, ,\, \frac{\6}{\6(\5\6\5\eta)}\,\Big\}w^I
=\5L_0\,w^I \ .
\label{CONTRACT}
\end{equation}
Hence, in the space of local functions of the $w$'s
the derivatives with respect to $\6\eta$ and $\5\6\5\eta$
are contracting homotopies for $L_0$ and $\5L_0$, respectively,
and the cohomology can be nontrivial only in the 
intersection of the kernels of $L_0$ and $\bar L_0$.

All $w$'s are eigenfunctions of $L_0$ and $\bar L_0$ with the 
eigenvalues being their ``conformal weights''. The only $w^I$ with 
negative conformal weights are the
undifferentiated diffeomorphism ghosts $\eta,\bar\eta$ and the 
undifferentiated supersymmetry ghosts $\ep,\bar\ep$;
their conformal weights are $(-1,0)$, $(0,-1)$, $(-1/2,0)$ and
$(0,-1/2)$, respectively [here $(a,b)$ are the eigenvalues
of $(L_0,\bar L_0)$]. The only
superconformal tensor fields with vanishing conformal weights
are the undifferentiated $X^M$. These properties
simplify the analysis enormously.

Our strategy for finding the solutions to (\ref{cc}) will be based on an 
expansion in supersymmetry ghosts
\begin{eqnarray}
\omega &=& \sum_{k=0}^{\bar k}\omega_k\ ,\quad
(N_{\ep}+N_{\bar\ep})\omega_k=k\omega_k
\nonumber\\
s&=& s_2+s_1+s_0\ ,\quad [N_\ep+N_{\bar \ep},s_k]=ks_k ,
\label{expansion}
\end{eqnarray}
where we have introduced the counting operator $N_{\ep}$ for the susy ghost 
$\ep$ and all its derivatives
\begin{equation}
N_{\ep}=\sum_{n\geq 0}(\partial^n\ep)\,
\frac{\partial}{\partial(\partial^n\ep)}
\label{Nep}
\end{equation}
and analogously $N_{\bar\ep}$ counts $\bar\ep$ and derivatives thereof.%
\footnote{We note that the expansion 
(\ref{expansion}) holds because we are studying the antifield
independent cohomology here. The analogous expansion
in presence of antifields is more involved; in fact, it can even
involve infinitely many terms. Therefore the strategy applied here
to determine the action is not practicable in the same way for analysing
the full (antifield dependent) cohomology later.}
One observes that $s_2$ is the simplest piece in the above decomposition 
of $s$. It acts nontrivially only on the reparametrization ghosts $\eta$, 
$\bar\eta$, derivatives thereof and on $C^i$,
\[
s_2\eta=-\ep\ep\ ,\quad s_2\bar\eta=-\bar\ep\bar\ep\ ,\quad
s_2C^i=\ep\bar\ep\hat\phi^i\ .
\]
We shall base the investigation on the cohomology of $s_2$. 
The cocycle condition $s\omega=0$ decomposes into
\begin{equation}
s_2\omega_{\bar k}=0,\quad s_1\omega_{\bar k}+s_2\omega_{\bar k-1}=0,
\quad \dots
\label{chain}
\end{equation}
Due to the requirement of ghost number $2$ and antifield number $0$ in 
(\ref{cc}), one is left with $0\leq\bar k\leq 2$. The three possible values 
for $\bar k$ are now analysed case by case.
\\
%
%
\underline{$\bar k$=0:} The general form of $\omega_{\bar 0}$ according 
to the condition of vanishing conformal weight is 
\begin{eqnarray*}
\omega_{\bar 0}&=& \eta\bar\eta A_{(1,1)}+\eta\partial\eta A_{(1,0)}
+\bar\eta\bar\partial\bar\eta A_{(0,1)}+\eta\bar\partial\bar\eta B_{(1,0)}
+\bar\eta\partial\eta B_{(0,1)}
\\
& &+\eta\partial^2\eta A_{(0,0)}+\bar\eta\bar\partial^2\bar\eta \bar A_{(0,0)}
+\partial\eta\bar\partial\bar\eta B_{(0,0)}+C^iC^j D_{ij(0,0)}
\\
& &+\eta C^i D_{i(1,0)}+\bar\eta C^i D_{i(0,1)}
+\partial\eta C^i D_{i(0,0)}+\bar\partial\bar\eta C^i \bar D_{i(0,0)},
\end{eqnarray*}
where the $A$'s, $B$'s and $D$'s do not depend on the ghosts and the 
subscripts $(m,n)$ indicate their conformal weights. It is easy to verify 
explicitly that
\begin{equation} 
s_2\omega_{\bar 0}=0\ \Leftrightarrow\ \omega_{\bar 0}=0.
\label{k0}
\end{equation}
%
%
\underline{$\bar k$=1.} The general form of $\omega_{\bar 1}$ is 
\begin{eqnarray*}
\omega_{\bar 1}&=& \eta\ep A_{(3/2,0)}+ \bar\eta\bar\ep A_{(0,3/2)}
+\eta\bar\ep A_{(1,1/2)}+\bar\eta \ep A_{(1/2,1)}
\\
& &+\eta\partial\ep A_{(1/2,0)}+\bar\eta\bar\partial\bar\ep A_{(0,1/2)}
+\ep\partial\eta B_{(1/2,0)}+\bar\ep\bar\partial\bar\eta B_{(0,1/2)}
\\
& &+\ep\bar\partial\bar\eta C_{(1/2,0)}+\bar\ep\partial\eta C_{(0,1/2)}
+\ep C^i D_{i(1/2,0)}+\bar\ep C^i D_{i(0,1/2)},
\end{eqnarray*}
where again the $A$'s, $B$'s and $D$'s do not depend on the ghosts and 
their conformal weights are indicated in brackets. A straightforward 
computation shows that $s_2\omega_{\bar 1}=0$ imposes
\begin{eqnarray*}
&  A_{(3/2,0)}=A_{(0,3/2)}=C_{(1/2,0)}=C_{(0,1/2)}=0 &
\\
&  A_{(1,1/2)}=\hat\phi^i D_{i(1/2,0)}\ ,\quad
   A_{(1/2,1)}=\hat\phi^i D_{i(0,1/2)} &
\\
&  A_{(1/2,0)}=-2B_{(1/2,0)}\ ,\quad
   A_{(0,1/2)}=-2B_{(0,1/2)} &
\end{eqnarray*}
The conformal weights $(1/2,0)$ and $(0,1/2)$ imply 
\begin{eqnarray*}
& D_{i(1/2,0)}=\psi^M D_{Mi}(X),\ D_{i(0,1/2)}=\bar\psi^M \bar D_{Mi}(X)&
\\
& B_{(1/2,0)}=\psi^M B_M(X),\ B_{(0,1/2)}=\bar\psi^M\bar B_M(X)&
\end{eqnarray*}
where we indicated that the remaining $B$'s and $D$'s are arbitrary 
functions of the $X$'s. Hence, we get
\begin{eqnarray*}
\omega_{\bar 1}&=&(\eta\bar\ep\hat\phi^i+\ep C^i)\psi^M D_{Mi}(X)
+(\bar\eta\ep\hat\phi^i+\bar\ep C^i)\bar\psi^M \bar D_{Mi}(X)
\\
& &+(\ep\partial\eta-2\eta\partial\ep)\psi^M B_M(X)
+(\bar\ep\bar\partial\bar\eta
-2\bar\eta\bar\partial\bar\ep)\bar\psi^M \bar B_M(X).
\end{eqnarray*}
The second equation (\ref{chain}) requires that $s_1\omega_{\bar 1}$
be $s_2$-exact. This imposes
\begin{eqnarray*}
& &
B_M=\bar B_M=0,\quad D_{Mi}=\bar D_{Mi}\ ,
\quad \partial_N\bar D_{iM}=\partial_M D_{iN}
\\
&\Leftrightarrow& B_M=\bar B_M=0,\quad D_{Mi}=\bar D_{Mi}=\partial_M D_i(X)
\end{eqnarray*}
where we have introduced the notation
\[
\partial_M=\frac{\partial}{\partial X^M}\ .
\]
Furthermore, the second equation (\ref{chain}) uniquely determines 
the function $\omega_0$, which corresponds to $\omega_{\bar 1}$ 
[the uniqueness follows from (\ref{k0})]. 
It turns out that the other equations (\ref{chain})
do not impose further conditions in this case, but are automatically
fulfilled. Altogether we find
\begin{eqnarray}
\omega_{\bar 1}&=&
[(\eta\bar\ep\hat\phi^i+\ep C^i)\psi^M
+(\bar\eta\ep\hat\phi^i+\bar\ep C^i)\bar\psi^M]\partial_M D_i(X)
\label{k11}\\
\omega_0&=&-\eta\bar\eta [\psi^M\bar\lambda^i-\bar\psi^M\lambda^i
+\hat F^M\hat\phi^i+\psi^M\bar\psi^N\hat\phi^i\partial_N]\partial_M D_i(X)
\nonumber\\
& &+C^i(\eta\mcd X^M+\bar\eta\bar\mcd X^M)\partial_M D_i(X)
\label{k10}
\end{eqnarray}
Using the freedom to add a coboundary we obtain by adding $s[C^iD_i(X)]$ 
to $\omega_{\bar 1}+\omega_0$ the equivalent solution 
\begin{eqnarray}
&\eta\bar\eta \cF^i D_i(X) 
-\eta\bar\eta(\psi^M\bar\lambda^i-\bar\psi^M\lambda^i
+\hat F^M\hat\phi^i+\psi^M\bar\psi^N\hat\phi^i\partial_N)\partial_M D_i(X)&
\nonumber\\
&+\eta\bar\ep(\lambda^i+\hat\phi^i\psi^M\partial_M) D_i(X)
+\bar\eta\ep(\bar\lambda^i+\hat\phi^i\bar\psi^M\partial_M) D_i(X)
+\ep\bar\ep\hat\phi^i D_i(X).
\label{k11a}
\end{eqnarray}
%
%
\underline{$\bar k$=2.} The general form of $\omega_{\bar 2}$ is given by 
\begin{eqnarray*}
\omega_{\bar 2}&=& \ep\ep A_{(1,0)}+\bar\ep\bar\ep A_{(0,1)}
+\ep\bar\ep A_{(1/2,1/2)}+\ep\partial\ep B(X)
+\bar\ep\bar\partial\bar\ep\bar B(X),
\end{eqnarray*}
where due to the indicated conformal weights one has  
\begin{eqnarray*}
 A_{(1,0)}&=& \mcd X^M A_M(X)+\psi^M\psi^N A_{MN}(X)
\\
 A_{(0,1)}&=& \bar\mcd X^M \bar A_M(X)+\bar\psi^M\bar\psi^N \bar A_{MN}(X)
\\
 A_{(1/2,1/2)}&=& \hat F^M H_M(X)+\hat\phi^i H_i(X)+\psi^M\bar\psi^N H_{MN}(X)
\end{eqnarray*}
We can simplify $\omega_{\bar 2}$ using the freedom to subtract $s$-exact 
pieces from an $s$-cocycle. In particular, we can therefore neglect pieces
in $\omega_{\bar 2}$ which are of the form
$s_1\hat\omega_1+s_2\hat\omega_0$ (i.e. we consider 
$\omega'=\omega-s(\hat\omega_1+\hat\omega_0)$ where
$\omega$ is an $s$-cocycle arising from  $\omega_{\bar 2}$). Choosing 
\[
\hat\omega_1=\sfrac 12\,(\bar\ep\bar\psi^M-\ep\psi^M) H_M(X)
\]
we get
\begin{eqnarray*}
s_1\hat\omega_1=\ep\bar\ep \hat F^M H_M(X)
+\sfrac 12(\bar\ep\bar\ep\bar\mcd X^M-\ep\ep\mcd X^M)H_M(X)
\\
-\sfrac 12(\bar\ep\bar\psi^M-\ep\psi^M)(\bar\ep\bar\psi^N+\ep\psi^N)
\partial_N H_M(X).
\end{eqnarray*}
This shows that by subtracting
$s_1\hat\omega_1$ from $\omega_{\bar 2}$, we can remove
the piece $\hat F^M H_M(X)$ from $A_{(1/2,1/2)}$, thereby
redefining $A_{(1,0)}$, $A_{(0,1)}$ and $H_{MN}(X)$.
Furthermore, we have
\begin{eqnarray*}
&
\ep\ep A_{(1,0)}+\bar\ep\bar\ep A_{(0,1)}
+\ep\bar\ep \hat\phi^i H_i(X)+\ep\partial\ep B(X)
+\bar\ep\bar\partial\bar\ep\bar B(X)
= s_2\hat\omega_0,&
\\
&\hat\omega_0=-\eta A_{(1,0)}-\bar\eta A_{(0,1)}
+C^i H_i(X)-\sfrac 12\partial\eta B(X)-\sfrac 12\bar\partial\bar\eta\bar B(X).&
\end{eqnarray*}
Hence, we can also remove the pieces containing
$A_{(1,0)}$, $A_{(0,1)}$, $H_i(X)$, $B(X)$ and $\bar B(X)$
from $\omega_{\bar 2}$. Without loss of generality,
we can thus restrict the investigation of the case $\bar k=2$ to
\begin{equation}
\omega_{\bar 2}=\ep\bar\ep\psi^M\bar\psi^N H_{MN}(X).
\label{k22}
\end{equation}
Obviously $\omega_{\bar 2}$ satisfies the first eqation (\ref{chain}),
since it does not involve $\eta$, $\bar\eta$ or $C^i$. One now has to analyze 
the remaining equations (\ref{chain}). It is straightforward to compute 
$s_1\omega_{\bar 2}$ and to verify that the second equation (\ref{chain}) is 
solved by
\begin{eqnarray}
\omega_1&=&\eta\bar\ep
[\mcd X^M\bar\psi^N-\psi^M\hat F^N+\psi^M\bar\psi^N\psi^K\partial_K]H_{MN}(X)
\nonumber\\
&&
+\bar\eta\ep
[-\psi^M\bar\mcd X^N-\hat F^M\bar\psi^N
+\psi^M\bar\psi^N\bar\psi^K\partial_K]H_{MN}(X).
\label{k21}
\end{eqnarray}
The third eq.\ (\ref{chain}) requires that
$s_0\omega_{\bar 2}+s_1\omega_1$ be $s_2$-exact. This turns out
to be the case (for arbitrary $H_{MN}$) and determines
$\omega_0$. One finds
\begin{eqnarray}
\omega_0&=&\eta\bar\eta \Omega,
\nonumber\\
\Omega&=&(\mcd X^M\bar\mcd X^N+\hat F^M\hat F^N+\mcd\bar\psi^M\bar\psi^N
         -\psi^M\bar\mcd\psi^N)H_{MN}(X)
\nonumber\\
& & -(\mcd X^M\bar\psi^N\bar\psi^K+\bar\mcd X^N\psi^M\psi^K)
       \partial_K H_{MN}(X)
\nonumber\\
& & +(\hat F^M\psi^K\bar\psi^N-\hat F^K\psi^M\bar\psi^N
       +\hat F^N\psi^M\bar\psi^K)\partial_K H_{MN}(X)
\nonumber\\
& &+\psi^M\psi^K\bar\psi^N\bar\psi^L\partial_K\partial_L H_{MN}(X).
\label{k20}
\end{eqnarray}
The remaining two equations (\ref{chain}) are also satisfied. 
The functions $H_{MN}(X)$ are completely arbitrary. The symmetrized part 
$H_{(MN)}(X)$ and  the antisymmetrized part $H_{[MN]}(X)$ give rise to the
``target space metric'' $G_{MN}$ and the ``Kalb-Ramond field'' $B_{MN}$, 
respectively. Despite of our string inspired terminology we stress that 
there are no conditions imposed on $G_{MN}$ and $B_{MN}$ apart from their 
symmetry properties. In particular the ``metric'' $G_{MN}$ need not be 
invertible (in section \ref{sigmacoho} we shall impose 
that a submatrix of $G_{MN}$ be invertible). 
$B_{MN}$ is determined only up to
\[
H_{[MN]}(X)\rightarrow H_{[MN]}(X)+\partial_{[M}B_{N]}(X)
\]
where $B_M(X)$ are arbitrary functions. This
originates from the fact that the $s$-cocycle
$\omega=\omega_{\bar 2}+\omega_1+\omega_0$ remains form invariant under
\[
\omega\rightarrow \omega+s[(\ep\psi^M+\bar\ep\bar\psi^M
+\eta\mcd X^M+\bar\eta\bar\mcd X^M)B_M(X)+\dots]
\]
where the dots stand for terms at least bilinear in the fermions.
Changing $\omega$ by such $s$-exact pieces results in the
above change of $H_{[MN]}(X)$ and modifies the Lagrangian
by a total derivative.
%
%
\subsection{Result}\label{result}
 
We conclude that up to redefinitions by coboundary terms, the general 
solution of (\ref{cc}) is given by the sum of the functions 
(\ref{k11a})--(\ref{k20}). The solution involves arbitrary
functions $D_i(X)$ and $H_{MN}(X)$, which thus parametrize
the various possible actions. The antisymmetric part of $H_{MN}(X)$ is 
determined only up to redefinitions by $\partial_{[M}B_{N]}(X)$, as
$H_{MN}(X)\rightarrow H_{MN}(X)+\partial_{[M}B_{N]}(X)$ modifies the
Lagrangian only by total derivatives. The functions $D_i(X)$ are
determined up to arbitrary constants, since only $\partial_M D_i(X)$ enters 
in the equivalent solution (\ref{k11}) and (\ref{k10}).\footnote{A 
constant in $D_i$ yields a topological term in the action 
proportional to the Chern class of the gauge bundle.}
Owing to general properties of descent equations in diffeomorphism 
invariant theories
\cite{Brandt:1990et,Barnich:1995ap,Dragon:1996md,Brandt:1997mh},
the integrand of the action is obtained from the
solution of (\ref{cc}) simply by substituting world-sheet
differentials for diffeomorphism ghosts $\xi^m$.
The resulting Lagrangian, written in
terms of the Beltrami fields, is a 
generalized version of the one found in~\cite{Delduc:1990gn}: 
\begin{eqnarray}
L &=&L_{Matter}+L_{U1}
\nn\\
L_{Matter} &=& \sfrac{1}{1-\m\bar\m}{\big [} 
       (\partial-\bar\m\bar\partial)X^M (\bar\partial-\m\partial)X^N (G_{MN} 
               + B_{MN}) \nn
\\
    & &  -\((\partial - \bar\m\bar\partial)X^M \a \ps^N  
         +(\bar\partial - \m\partial)X^M \bar\a \bar \ps^N\) G_{MN} 
         -\sfrac{1}{2} \a\bar\a \ps^M \bar\ps^N G_{MN}{\big ]} \nn
\\ 
    & &  -\(\bar\ps^N(\partial - \bar\m\bar\partial)\bar\ps^M   
         +\ps^N(\bar\partial - \m\partial)\ps^M\)G_{MN}
         -(1-\m\bar\m) \hat F^M \hat F^N G_{MN}\nn
\\
    & &  -\bar\ps^M \bar\ps^N(\partial-\bar\m\bar\partial)X^K 
          (\G_{KNM}-\sfrac{1}{2}H_{KNM}) \nn
\\ 
    & &  -\ps^M\ps^N(\bar\partial-\m\partial)X^K 
          (\G_{KNM}+\sfrac{1}{2}H_{KNM})\nn
\\ 
    & &  +\sfrac{1}{6}(\bar\a\bar\ps^M\bar\ps^N\bar\ps^K-\a\ps^M\ps^N\ps^K) 
                H_{KMN} \nn
\\
    & &  +(1-\m\bar\m)\hat F^M\ps^K\bar\ps^N(2 \G_{KNM}-H_{KNM})  \nn
\\ 
    & &  +\sfrac{1}{2}(1-\m\bar\m)\ps^M\ps^K\bar\ps^N\bar\ps^L R_{KMLN} \nn
\\
L_{U1} &=& F^i D_i - (1-\m\bar\m)[\ps^M(\bar \l^i 
       -\sfrac{1}{2} \sfrac{1}{1-\m\bar\m} \m\bar\a \hat\Ph^i)
       -\bar\ps^M(\l^i-\sfrac{1}{2}\sfrac{1}{1-\m\bar\m}\bar\m\a\hat\Ph^i) \nn 
\\
      & &  +\hat F^M\hat\Ph^i+\ps^M\bar\ps^N\partial_N]\partial_M D_i 
\label{action1}
\end{eqnarray}
where we have introduced the following notations
\begin{eqnarray}
G_{MN} & := &  H_{(MN)}(X) \qquad B_{MN} := H_{[MN]}(X) \nn
\\
D_i &:=& D_i(X) \qquad F^i:=\e^{mn}(\partial_m A^i_n-\partial_n A^i_m) \nn
\\
\O_{KNM} &:=& \partial_K H_{MN}(X)-\partial_M H_{KN}(X)+\partial_N H_{KM}(X) 
            = 2\G_{KNM} - H_{KNM} \nn 
\\
R_{KLMN} &:=& \partial_M\partial_{[K}H_{L]N}(X)
              -\partial_N\partial_{[K}H_{L]M}(X) \nn
\end{eqnarray}
The ``target space curvature'' $R_{KLMN}$ we have 
introduced is of course not the 
Riemannian one. The Riemannian curvature appears after eliminating the 
auxiliary fields from the action. 

Of course, the
action can be also written in terms of the original fields introduced 
in section 
\ref{fieldcont}. One obtains from the matter part the well known superstring 
action including the B-field background~\cite{Bergshoeff:1986qr} 
\begin{eqnarray}
L/e&=& \mbox{$\frac {1}{2}$}
       \partial_m X^M\partial_n X^N (-h^{mn}G_{MN}+\varepsilon^{mn}B_{MN}) 
       +\mbox{$\frac {{\mathrm{i}}}{2}$} 
               \overline\psi^M\gamma^m\partial_m\psi^N G_{MN} \nn
\\
   & & +\mbox{$\frac {1}{2}$} F^M F^N G_{MN} \nonumber
       +\chi_k\gamma^n \gamma^k (\psi^N\partial_n X^M
       -\mbox{$\frac {1}{4}$} C\chi_n \overline\psi^M\psi^N)G_{MN} \nonumber
\\  
     & &  + (\mbox{$\frac {1}{2}$} F^M \overline\psi^K \psi^N  
         -{{\mathrm{i}}}\,\overline\psi^N\gamma^m\psi^M \partial_m X^K) 
         \Gamma_{NKM} \nonumber
\\
     & &  +\mbox{$\frac {1}{4}$}(F^M \overline\psi^K \gamma_* \psi^N
          - {\mathrm{i}} \overline\psi^N\gamma^m\gamma_*\psi^M \partial_m X^K)
          H_{NKM} \nonumber
\\
     & &  - \mbox{$\frac {{\mathrm{i}}}{12}$} 
          \chi_m\gamma^n\gamma^m\psi^M\overline\psi^N\gamma_n \gamma_*\psi^K 
          H_{MNK} \nonumber
\\
     & &  +\mbox{$\frac {1}{16}$}
          \overline\psi^M({1\hspace{-0.243em}\mathrm{l}} + \gamma_*)
          \psi^N\overline\psi^K({1\hspace{-0.243em}\mathrm{l}}
          +\gamma_*)\psi^L R_{KMLN}
          \nonumber
\\
     & & + \varepsilon^{mn} D_i\partial_m A_n^i 
         +\mbox{$\frac {{\mathrm{i}}}{4}$}\overline\psi^M\psi^N \phi^i 
         \partial_N\partial_M D_i  \nonumber
\\
       & & + \mbox{$\frac {1}{2}$}({\mathrm{i}} \overline\psi^N
           \gamma_* \lambda^i  
           -{\mathrm{i}} F^N \phi^i+ \chi_m \gamma^m\psi^N \phi^i) 
           \partial_N D_i .
\label{action2}
\end{eqnarray}
Thus the cohomological analysis shows that in the absence of gauge 
multiplets the Lagrangian derived in~\cite{Bergshoeff:1986qr} is in fact 
unique up to total derivatives and choices of the background fields. 
It should be kept in mind, however, that this uniqueness is tied to the gauge 
transformations 
specified in section~\ref{fieldcont}. It can get lost when one allows
that the gauge transformations get deformed. This deformation
problem can be analysed by BRST cohomological means too, but then
the relevant cohomological problem includes the 
antifields \cite{Barnich:1993vg}. The results which we 
shall derive in the
second part of this work imply that the nontrivial consistent deformations
correspond one-to-one
to the deformations of the bosonic string models.
All deformations of bosonic string models
without world-sheet gauge fields
were derived in \cite{Brandt:1996nn}.
Hence, all nontrivial
deformations of the standard superstring
world-sheet action \cite{Bergshoeff:1986qr} and its gauge transformations
are supersymmetric generalizations of the actions and gauge
transformations given in \cite{Brandt:1996nn}; in particular
this result implies that nontrivial deformations
of the standard superstring gauge transformations exist 
only if the background possesses special Killing vectors described
in \cite{Brandt:1996nn}.
A full analysis (to all orders
in the deformation parameters) of the deformation problem 
for bosonic
models with world-sheet gauge fields is missing so far,
but a complete classification of
the first order deformations was given in \cite{Brandt:1998cy}. 
The latter results
extend thus to the superstring models too.

\mysection{Antifields}\label{antifields}

To proceed with our analysis we have to bring the antifields  
into the game. According to the principles of the field-antifield 
formalism~\cite{Batalin:1981jr,Batalin:1983jr,Henneaux:1992ig,Gomis:1995he} 
to each field a corresponding antifield $\PH_A^*$ is introduced with 
ghost number and statistics
\[
\gh(\PH_A^*)=-\gh(\PH^A)-1, \quad 
\epsilon(\PH_A^*)=\epsilon(\PH^A)+1 ~\textrm{(mod 2)},
\]
such that the statistics of the antifields is opposite to that of the 
corresponding fields. It is useful to introduce still another grading into the 
algebra of fields and antifields, namely the already mentioned antifield 
(or antighost) number. On all the fields (including the ghosts) the antifield 
number is defined to be zero, i.e., $\agh(\PH^A)=0$. On the antifields the 
antifield number equals minus the ghost number, 
$\agh(\PH_A^*)=-\gh(\PH_A^*)$.

The antibracket for two arbitrary functions of the fields $\PH^A$ and 
antifields $\PH_A^*$ is defined as 
\[
(F,G)=\int \(\frac{\d_R F}{\d \PH^A}\frac{\d_L G}{\d \PH_A^*} - 
      \frac{\d_R F}{\d \PH_A^*}\frac{\d_L G}{\d \PH^A}\).
\]
Thus the antibracket has odd statistics and carries 
ghost number one. The BRST transformations of the antifields are 
generated via the antibracket by the proper solution $\cS$ to the classical 
master equation $(\cS,\cS)=0$ according to 
\[
s \PH_A^* = (\cS,\PH_A^*) = \frac{\d_R \cS}{\d \PH^A}.
\]
Owing to the off-shell closure of the gauge algebra $\cS$
simply reads
\[
\cS = S_0 - \int (s \PH^A)\PH_A^* ,
\] 
where $S_0$ is the classical action and
$s \PH^A$ are the BRST transformations given in
section \ref{fieldcont}. It is useful to decompose the BRST 
differential according to the grading with respect to the antifield 
number $s=\sum_{k\ge -1}s_k$ with ${\agh}(s_k)=k$
(this decomposition should not be confused with the one
in (\ref{expansion}) even though we use the same notation). 
The decomposition starts 
with the field theoretical Koszul-Tate differential 
$\delta\equiv s_{-1}$ and the differential $\g\equiv s_0$. 
Contrary to the bosonic case the decomposition does not terminate at this 
level. An additional part $s_1$ raising the antifield number by one unit 
shows up reflecting field dependent gauge transformations in the commutator
of supersymmetry transformations. The Koszul-Tate 
differential acts nontrivially only on the antifields and implements the
equations of motion. Hence, the knowlegde of the classical action is necessary 
to determine the $\delta$-transformations of the antifields. However, the 
action of the part of the BRST differential leaving the antifield number 
unchanged is determined solely by the imposed gauge transformations. The 
$\g$-transformations of the antifields corresponding to the 
matter fields and the $U(1)$ multiplet read 
\begin{eqnarray}\label{santifields}
\g X^*_M &=&\partial_m(\x^m X^*_M)  
            -\Ii\partial_m(\x^{\b}(\g^mC)_{\b\a}{}\ps_M^{*\a})\nn
\\
       & &  -\sfrac{1}{2} \partial_m(\x^{\a}(\g^n\g^mC)_{\a\b}
             \chi_n^{\b}F^*_M) \nn
\\
\g \ps^{*\a}_M &=& \partial_m(\x^m \ps^{*\a}_M)+\x^{\a}X^*_M 
                -\Ii\x^{\g}(\g^mC)_{\g\b}\chi_m^{\a}\ps^{*\b}_M 
                -\sfrac{\Ii}{2}\partial_m(\x^{\b}(\g^m)_{\b}^{~\a}F^*_M) \nn 
\\
            & & -\sfrac{\Ii}{8}\x^{\b}(\g^m\g_*)_{\b}^{~\a}
                           \o_m^{~ab}\e_{ab}F^*_M 
                -\sfrac{1}{2}\x^{\b}\chi_m^{~\d}(\g^m\g^nC)_{\b\d}
                           \chi_n^{~\a}F^*_M \nn
\\
         & & -\sfrac{1}{4}C^{ab}\e_{ab}(\g_*)_{\b}^{~\a}\ps^{*\b}_M 
             +\sfrac{1}{2} C^W \ps^{*\a}_M \nn
\\
\g F^*_M &=&  \partial_m(\x^m F^*_M)
            -\x^{\b}C_{\b\a}\ps^{*\a}_M 
            -\sfrac{\Ii}{2}\x^{\b}(\g^mC)_{\b\a}\chi_m^{\a}F^*_M 
            +C^W F^*_M \nn
\\
\g A^{*m}_i &=& \partial_n(\x^n A^{*m}_i)
             -(\partial_n\x^m)A^{*n}_i \nn
\\
 & &         +\Ii\partial_n(\x^{\a}(\g_*C)_{\a\b}\e^{nm}\l_i^{*\b}) \nn
\\
\g \Ph^*_i &=& \partial_m(\x^m \Ph^*_i)
            -\x^\a(\g_*C)_{\a\b}\ep^{mn}\chi_n(\g_*C)\chi_m\l_i^{*\b}\nn
\\
        & & -\Ii\partial_m(\x^\a(\g_*\g^mC)_{\a\b}\l_i^{*\b}
            -\sfrac{\Ii}{2}\x^{\a}(\g_*C)_{\a\b}S\l^{*\b}_i \nn
\\
        & & +2\Ii \x^{\a}\chi_m^{\b}(\g_*C)_{\b\a}A^{*m}_i
            -2\h^{\g}(\g_*C)_{\g\b}\l^{*\b}_i
            +C^W\Ph^*_i \nn
\\
\g \l^{*\a}_i &=& \partial_m(\x^m \l^{*\a}_i) 
               -\x^{\b}(\g_m)_{\b}^{~\a}A^{*m}_i 
               +\x^{\b}(\g_*)_{\b}^{~\a}\Ph^*_i 
               -\Ii\x^\b(\g_*\g^mC)_{\b\g}(\chi\g_*)^\a\l_i^{*\g}\nn 
\\
            & & -\Ii \x^{\d}(\g_*C)_{\d\b}\e^{kl}
                           \(\chi_k^{\g}(\g_l)_{\g}^{~\a}\)\l^{*\b}_i 
                -\sfrac{1}{4}C^{ab}\e_{ab}(\g_*)_{\b}^{~\a}\l^{*\b}_i
                +\sfrac{3}{2}C^W \l^{*\a}_i .
\end{eqnarray}
$s_1$ acts nontrivially on $A^{*m}_i$, $\Ph^*_i$ and on the antifields 
for the gravitational multiplet $\chi^{*m}_\alpha$, $e^{*m}_a$ and $S^*$. 
In particular one finds
\begin{equation}
s_1 A^{*m}_i = \Ii\x^{\a}\x^{\b}(\g^mC)_{\b\a} c^*_i, \qquad
s_1 \Ph^*_i = - \Ii \x^{\a}\x^{\b}(\g_*C)_{\b\a} c^*_i, \nn
\end{equation}
where $c^*_i$ denote the antifields corresponding to $U(1)$ ghosts.

The explicit form of the BRST transformations of the antifields for the 
gravitational multiplet and the ghosts will not be needed in the following. 
In section \ref{ISO} it is shown that they do not contribute 
nontrivially to the cohomology, at least at ghost number $g<4$.

\subsection{Superconformal antifields}\label{covantifields}

We shall now identify ``superconformal antifields''
whose $\gamma$-transformations take the same form
as the $s$-transformations of superconformal
tensor fields in (\ref{tensortrafo}).
The identification of superconformal antifields is somewhat more involved than
the procedure for the fields. From experience with the bosonic 
case one expects reasonable candidates to arise from redefinitions 
of the form $\Phi^*_A \rightarrow \sfrac{1}{1-\mu\bar\mu}\Phi^*_A$, 
accounting for the fact that antifields 
transform under diffeomorphisms as tensor densities rather than tensors. 
In addition we have to take care of their ``structure group 
transformations'', i.e., of their conformal weights, their Lorentz 
transformations and super-Weyl transformations\footnote{Antifields transform 
``contragradiently'' under structure group transformations 
as compared to the corresponding fields.}. Yet this does not suffice to obtain 
$\g$-transformations of the desired form. 
It turns out that the antifields have to be mixed among themselves. 
These considerations lead us to the following definitions of the lowest 
order matter antifields
\begin{eqnarray}
\hat F^*_M &\equiv& F^*_{M(0,0)}= \frac{1}{1-\m\bar\m}
               (e_{z}^{~z}e_{\bar z}^{~\bar z})^{-\sfrac{1}{2}}F^*_M \nn
\\
\hat\ps^*_M &\equiv& \ps^*_{M(0,0)}= \frac{\Ii}{\sqrt{2}}\frac{1}{1-\m\bar\m}
                (e_{z}^{~z})^{-\sfrac{1}{2}}\ps^*_M{}^{2}
                +\frac{\m\bar\a}{1-\m\bar\m}\hat F^*_M  \nn
\\
\hat{\bar\ps}{}^*_M & \equiv & \bar\ps^*_{M(0,0)}
                       =\frac{\Ii}{\sqrt{2}}\frac{1}{1-\m\bar\m}
                        (e_{\bar z}^{~\bar z})^{-\sfrac{1}{2}}\ps^*_M{}^{1}
                        -\frac{\bar\m\a}{1-\m\bar\m}\hat F^*_M \nn
\\
\hat X^*_M &\equiv& X^*_{M(0,0)} = \frac{1}{1-\m\bar\m} X^*_M
               +\frac{\bar\m\a}{1-\m\bar\m}\hat\ps^*_M 
               +\frac{\m\bar\a}{1-\m\bar\m}\hat{\bar\ps}{}^*_M
               +\frac{\a\bar\a}{1-\m\bar\m}\hat F^*_M. \nn
\end{eqnarray}
Their $\g$-transformations are indeed of the desired form (\ref{tensortrafo})
and read explicitly
\begin{eqnarray}
  \label{lowantifs}
  \g\hat F^*_M &=& 
      (\h\mathcal{D}+{\bar\h}\bar\mathcal{D})\hat F^*_M 
      -\e\hat{\bar\ps}{}^*_M + {\bar\e}\hat\ps^*_M 
      +\sfrac{1}{2}((\partial \h)+(\bar\partial{\bar\h}))\hat F^*_M \nn
\\
\g\hat\ps^*_M &=& 
               (\h\mathcal{D}+{\bar\h}\bar\mathcal{D})\hat\ps^*_M 
               +\e\hat X^*_M + {\bar\e}\bar\mathcal{D}\hat F^*_M 
               +(\sfrac{1}{2}(\partial \h) + (\bar\partial\bar\h))\hat\ps^*_M 
               +(\bar\partial{\bar\e})\hat F^*_M \nn 
\\
\g\hat{\bar\ps}{}^*_M &=& 
  (\h\mathcal{D}+{\bar\h}\bar\mathcal{D})\hat{\bar\ps}{}^*_M 
  +{\bar\e}\hat X^*_M - \e\mathcal{D}\hat F^*_M 
  +((\partial\h)+\sfrac{1}{2}(\bar\partial{\bar\h}))
     \hat{\bar\ps}{}^*_M 
  -(\partial\e)\hat F^*_M \nn
\\
\g\hat X^*_M &=& 
  (\h \mathcal{D}+{\bar\h}\bar\mathcal{D})\hat X^*_M 
  +\e\mathcal{D}\hat\ps^*_M 
  +{\bar\e}\bar\mathcal{D}\hat{\bar\ps}{}^*_M 
  +((\partial\h) + (\bar\partial{\bar\h}))\hat X^*_M \nn
\\ 
  & & +(\partial\e)\hat\ps^*_M 
      +(\bar\partial{\bar\e})\hat{\bar\ps}{}^*_M .
\end{eqnarray}
The expressions above are in fact already complete, since $s_1$ does not act 
nontrivially on the matter antifields. Analogously to the situation of the 
superconformal tensor fields the algebra (\ref{suvirasoro}) is represented 
on these fields and their derivatives, which we denote by 
\[
F^*_{M(m,n)}=(L_{-1})^m(\bar L_{-1})^n \hat F^*_M 
\equiv (\cD)^m(\bar\cD)^n \hat F^*_M, 
\]
etc, where the operators $L_{-1}$ and $\bar L_{-1}$ are identified with 
supercovariant derivatives as in (\ref{covdevs}). In particular one finds 
on the antifields with lowest conformal weights the following expressions 
\begin{eqnarray}
\mathcal{D}\hat F^*_M &=& \sfrac{1}{1-\m\bar\m}
       \((\partial - \bar\m\bar\partial 
      -\sfrac{1}{2} (\bar\partial\bar\m) 
      +\sfrac{1}{2} \bar\m (\partial\m))\hat F^*_M 
      -\sfrac{1}{2}\bar\m\a\hat{\bar\ps}{}^*_M 
      -\sfrac{1}{2}\bar\a\hat\ps^*_M \) \nn 
\\
\bar\mathcal{D}\hat F^*_M &=& \sfrac{1}{1-\m\bar\m}
                       \((\bar\partial - \m\partial 
                      -\sfrac{1}{2}(\partial\m) 
                      +\sfrac{1}{2} \m (\bar\partial\bar\m))\hat F^*_M 
                      +\sfrac{1}{2}\a\hat{\bar\ps}{}^*_M 
                      +\sfrac{1}{2} \m\bar\a\hat\ps^*_M \) \nn
\\
\mathcal{D}\hat\ps^*_M &=& \sfrac{1}{1-\m\bar\m}
  \((\partial-\bar\m\bar\partial
       -(\bar\partial\bar\m)+\sfrac{1}{2}\bar\m(\partial\m))\hat\ps^*_M 
    -\sfrac{1}{2}\bar\m\a\hat X^*_M
    -\sfrac{1}{2}\bar\a\bar\mathcal{D}\hat F^*_M 
    -\sfrac{1}{2}(\bar\partial\bar\a)\hat F^*_M\) \nn
\\
\bar\mathcal{D}\hat\ps^*_M &=& \sfrac{1}{1-\m\bar\m}
   \((\bar\partial-\m\partial
       -\sfrac{1}{2}(\partial\m)+\m(\bar\partial\bar\m))\hat\ps^*_M 
   +\sfrac{1}{2}\a\hat X^*_M 
   +\sfrac{1}{2}\bar\m\a\mathcal{D}\hat F^*_M
   +\sfrac{1}{2}\bar\m(\partial\a)\hat F^*_M\) \nn
\end{eqnarray}
and analogous formulas for $\mathcal{D}\hat{\bar\ps}{}^*_M$ and 
$\bar\mathcal{D}\hat{\bar\ps}{}^*_M$. Again higher order antifields will 
not be needed. 

The construction of the covariant antifields for the gauge multiplet 
follows the arguments given above, with the additional task to get rid of the 
super-Weyl transformations. We introduce the redefinitions 
\begin{eqnarray}
\hat\l^*_i &\equiv& \l^*_{i(0,0)} = -\frac{1}{1-\m\bar\m}
           (e_{\bar z}^{~\bar z})^{-\sfrac{1}{2}}(e_{z}^{~z})^{-1}\l^{*2} \nn
\\
\hat{\bar\l}{}^*_i &\equiv& \bar\l^*_{i(0,0)} = \frac{1}{1-\m\bar\m}
           (e_{z}^{~z})^{-\sfrac{1}{2}}(e_{\bar z}^{~\bar z})^{-1}\l^{*1} \nn
\\
\hat\Ph^*_i &\equiv& \Ph^*_{i(0,0)} = \frac{1}{\sqrt{2}}\frac{1}{1-\m\bar\m}
 (e_{z}^{~z})^{-\sfrac{1}{2}}(e_{\bar z}^{~\bar z})^{-\sfrac{1}{2}}\Ph^*_i \nn 
\\
  & &  -\frac{1}{2}\frac{1}{1-\m\bar\m}
          \(\hat\chi_z^{~2}-\bar\m\hat\chi_{\bar z}^{~2}\)\hat\l^*_i 
       -\frac{1}{2}\frac{1}{1-\m\bar\m}
          \(\hat\chi_{\bar z}^{~1}-\m\hat\chi_z^{~1}\)\hat{\bar\l}{}^*_i \nn
\\
\hat A^*_i &\equiv& A^*_{i(0,0)} = 
     \frac{1}{\sqrt{2}}\frac{1}{1-\m\bar\m}\(\bar A^*_i+\bar\m A^*_i\) 
     -\frac{1}{1-\m\bar\m}\(\bar\a\hat\l^*_i+\bar\m\a\hat{\bar\l}{}^*_i\) \nn
\\
\hat{\bar A}{}^*_i &\equiv& \bar A^*_{i(0,0)} = 
     \frac{1}{\sqrt{2}}\frac{1}{1-\m\bar\m}\(A^*_i+\m \bar A^*_i\) 
     -\frac{1}{1-\m\bar\m}\(\a\hat{\bar\l}{}^*_i+\m\bar\a\hat\l^*_i\), 
\end{eqnarray}
where we have used the shorthand notation for the corrections involving 
gravitions $\hat\chi_{z}^{~1}=\sqrt{\sfrac{8}{e_{\bar z}^{~\bar z}}}
\chi_{z}^{~1}$ and $\hat\chi_{z}^{~2}=\sqrt{\sfrac{8}{e_{z}^{~z}}}
\chi_{z}^{~2}$ with obvious expressions for the $\bar z$ components. 
The $\g$-transformations then read
\begin{eqnarray}
\g\hat\l^*_i &=& (\eta\cD+{\bar\eta}\bar\cD)\hat\l^*_i
             +\sfrac{1}{2}\bar\partial{\bar\eta}\lambda^*_i
             +\ep\hat\phi^*_i-{\bar\ep}\hat{\bar A}{}^*_i \nn
\\
\g\hat{\bar\l}{}^*_i &=& 
             (\eta\cD+{\bar\eta}\bar\cD)\hat{\bar\lambda}{}^*_i
             +\sfrac{1}{2}\partial\eta\hat{\bar\lambda}{}^*_i
             +{\bar\ep}\hat\phi^*_i-\ep \hat A^*_i \nn
\\
\g\hat\phi^*_i &=& (\eta\cD+{\bar\eta}\bar\cD)\hat\Ph^*_i 
     +\sfrac{1}{2}(\partial\eta+\bar\partial{\bar\eta})\hat\Ph^*_i
     +\ep\cD\hat\l^*_i+{\bar\ep}\bar\cD\hat{\bar\l}{}^*_i \nn
\\
\g \hat A^*_i  &=& 
     (\eta\cD+{\bar\eta}\bar\cD)\hat A^*_i
     +\partial\eta \hat A^*_i
     +{\bar\ep}\cD\hat\l^*_i-\ep\cD\hat{\bar\l}{}^*_i
     -\partial\ep\hat{\bar\lambda}{}^*_i \nn
\\
\g\hat{\bar A}{}^*_i &=& 
     (\eta\cD+{\bar\eta}\bar\cD)\hat{\bar A}{}^*_i
     +\bar\partial{\bar\eta}\hat{\bar A}{}^*_i
     +\ep\bar\cD\hat{\bar\l}{}^*_i-{\bar\ep}\bar\cD\hat\l^*_i
     -\bar\partial{\bar\ep}\hat\lambda^*_i,
\end{eqnarray}
and are indeed of the desired form respecting the requirement (\ref{uvw}).
Note that the combination of the gravitinos used in the redefinition of 
$\hat\Ph^*_i$ transforms into the super-Weyl ghost thereby removing the 
unwanted transformation properties under the super-Weyl symmetry. 
Again higher order antifields will not be needed.

The explicit form of the superconformal antifields given above has 
already been used to derive the results for the rigid symmetries presented 
in~\cite{Brandt:2000ib}. A complete list of the BRST transformations 
(including the Koszul-Tate part and the $s_1$-transformations) of the 
antifields needed for the cohomological analysis is given in appendix 
\ref{transformations2}. In the following sections (and also in the 
appendices) we have dropped the hats on the superconformal antifields, but it 
is clear from the context which set of variables is meant.

\mysection{On-shell cohomology}\label{sigmacoho}

We shall now define and analyse an
``on-shell BRST cohomology'' $H(\sigma)$
and show that it is isomorphic to its purely bosonic
counterpart at ghost numbers $<4$, i.e., to the on-shell BRST cohomology 
of the corresponding bosonic string model.
The relevance of $H(\sigma)$ rests on the fact that it is
isomorphic to the full local $s$-cohomology
$H(s)$ (in the jet space associated to the
fields and antifields), at least at ghost numbers $<4$,
\begin{equation}
g<4:\quad H^g(\sigma)\simeq H^g(s).
\label{iso}
\end{equation}
This will be proved in section \ref{ISO}.

The analysis in this and the next section is general, 
i.e., it applies to
any model with an action (\ref{action1}) (or,
equivalently,  (\ref{action2})) provided that 
two rather mild assumptions hold, which are introduced now.
The first assumption only 
simplifies the action a little bit but does not
reduce its generality:
as we have argued already in \cite{Brandt:2000ib},
one may assume that the functions $D_i(X)$ 
which occur in the action coincide
with a subset of the fields $X^M$. We denote this subset by
$\{y^i\}$ and the remaining $X$'s by $x^\mu$,
\begin{equation}
\{X^M\}=\{x^\mu,y^i\},\quad D_i(X)\equiv y^i.
\label{simple1}
\end{equation}
For physical applications this ``assumption'' does not 
represent any loss of generality because
it can always be achieved
by a field redefinition (``target space coordinate transformation'')
$X^M\rightarrow\tilde X^M=\tilde X^M(X)$. The $y^i$
may be interpreted as coordinates
of an enlarged target space leading to
``frozen extra dimensions'' \cite{Brandt:2000ib}.
The second assumption is  that $G_{\mu\nu}(x,y)$ is invertible
(in contrast, $G_{MN}$ need not be invertible).
This is particularly natural in the string theory context, 
since it allows one to interpret $G_{\mu\nu}$ as a target
space metric. It is rather likely that our result holds for even
weaker assumptions (but we did not study this question), 
because the results derived in \cite{Brandt:1996gu,Brandt:1998cy} 
for bosonic string models do not
use the invertibility of $G_{\mu\nu}$.

Let us remark that 
the isomorphism (\ref{iso}) is not too surprising, because it is
reminiscent of a standard
result of local BRST cohomology stating that
$H(s)$ is isomorphic to the on-shell cohomology of $\gamma$ in the
space of antifield independent functions, where $\gamma$ is
the part of $s$ with antifield number 0 (see, e.g., section 7.2 of
\cite{Barnich:2000zw}). However, 
(\ref{iso}) is not quite the same statement because the 
definition of $\sigma$ given below does not take the
equations of motion for $\mu$, $\5\mu$, $\alpha$ or $\5\alpha$
into account. Hence, (\ref{iso}) contains information
in addition to the standard
result of local BRST cohomology mentioned before:
the equations of motion for $\mu$, $\5\mu$, $\alpha$, $\5\alpha$
are not relevant to the cohomology! This is a useful
result as 
these equations of motion are somewhat unpleasant, 
because they are not linearizable
(the models under study do not fulfill the
standard regularity conditions
described, e.g., in section 5.1 of  \cite{Barnich:2000zw}).

\subsection{Definition of $\sigma$ and $H(\sigma)$}
\label{Hsigmadef}
 
$\sigma$ is an ``on-shell version'' of $s$ defined
in the space of local functions made of the
fields only (but not of any antifields). We work in the
`Beltrami basis' and use the equations of
motion obtained by varying the action (\ref{action1}) with respect
to the fields $X$, $\psi$, $\5\psi$, $\hat F$, $\hat \phi$, $\lambda$, 
$\5\lambda$ and $A_m$.
The covariant version of
these equations of motion can be obtained from
the $s$-transformations of the corresponding 
covariant antifields given in 
appendix \ref{transformations2} by setting the
antifield independent part (`Koszul-Tate part')
of these transformations to zero.
This gives the
following ``on-shell equalities'' ($\approx$):
\bea
\hat F^i&\approx&0
\label{Sigma1}\\
\psi^i&\approx&0
\label{Sigma2}\\
\5\psi^i&\approx&0
\label{Sigma3}\\ 
\cD y^i&\approx&0
\label{Sigma4}\\
\5\cD y^i&\approx&0
\label{Sigma5}\\
\hat\phi^i&\approx&2G_{i\mu}\hat F^\mu+\psi^\mu\5\psi^\nu\Omega_{\mu\nu i}
\label{Sigma6}\\
\lambda^i&\approx&2G_{i\mu}\cD\5\psi^\mu
+\cD x^\mu\5\psi^\nu\Omega_{\mu\nu i}
+\hat F^\mu\psi^\nu\Omega_{\nu i\mu}
+\psi^\mu\psi^\nu\5\psi^\rho R_{\nu\mu i\rho}
\label{Sigma7}\\
\5\lambda^i&\approx&-2G_{i\mu}\5\cD\psi^\mu
-\5\cD x^\mu\psi^\nu\Omega_{\nu\mu i}
+\hat F^\mu\5\psi^\nu\Omega_{i\nu\mu}
+\psi^\mu\5\psi^\nu\5\psi^\rho R_{\mu i\rho\nu}
\label{Sigma8}\\
\hat F^\rho &\approx&-\sfrac 12 \psi^\mu\5\psi^\nu{\Omega_{\mu\nu}}^\rho
\label{Sigma9}\\
\5\cD\psi^\mu&\approx&-\sfrac 12[\5\cD x^\nu\psi^\rho{\Omega_{\rho\nu}}^\mu
             +\sfrac 12 \psi^\lambda\5\psi^\sigma\5\psi^\rho
             {\Omega_{\lambda\sigma}}^\nu{\Omega^\mu}_{\rho\nu}
             +\psi^\nu\5\psi^\rho\5\psi^\sigma {R^\mu}_{\nu\sigma\rho}]
\label{Sigma10}\\
\cD\5\psi^\mu&\approx&\sfrac 12[-\cD x^\nu\5\psi^\rho{\Omega_{\nu\rho}}^\mu
             +\sfrac 12 \psi^\lambda\5\psi^\sigma\psi^\rho
             {\Omega_{\lambda\sigma}}^\nu{\Omega_\rho}^\mu{}_\nu
             +\psi^\nu\psi^\rho\5\psi^\sigma {R_{\rho\nu\sigma}}^\mu]
\label{Sigma11}\\
\cF^i&\approx&2G_{i\mu}\cD\5\cD x^\mu+
\cD x^\mu\5\cD x^\nu\Omega_{\mu\nu i}
             -\hat F^\mu \hat F^\nu\Omega_{i\mu\nu}
\nonumber\\
         & & -\cD\5\psi^\mu\5\psi^\nu\Omega_{i\nu\mu}
             +\psi^\mu\5\cD\psi^\nu\Omega_{\mu i\nu}
\nonumber\\
         & & -\cD x^\mu\5\psi^\nu\5\psi^\rho R_{\mu i\rho\nu}
             -\5\cD x^\mu\psi^\nu\psi^\rho R_{\rho\nu\mu i}
\nonumber\\
         & & -\hat F^\mu\psi^\nu\5\psi^\rho\6_i\Omega_{\nu\rho\mu}
             -\sfrac 12\psi^\mu\psi^\nu\5\psi^\rho\5\psi^\sigma
             \6_i R_{\nu\mu \sigma\rho}
\label{Sigma12}\\
\cD\5\cD x^\mu&\approx&\sfrac 12[-\cD x^\nu\5\cD x^\rho{\Omega_{\nu\rho}}^\mu
             +\hat F^\nu \hat F^\rho{\Omega^\mu}_{\nu\rho}
\nonumber\\
         & & +\cD\5\psi^\nu\5\psi^\rho{\Omega^\mu}_{\rho\nu}
             -\psi^\nu\5\cD\psi^\rho{\Omega_\nu}^\mu{}_\rho
\nonumber\\
         & & -\cD x^\sigma\5\psi^\nu\5\psi^\rho {R^\mu}_{\sigma\rho\nu}
             +\5\cD x^\sigma\psi^\nu\psi^\rho {R_{\rho\nu\sigma}}^\mu
\nonumber\\
         & & +\hat F^\sigma\psi^\nu\5\psi^\rho\6^\mu\Omega_{\nu\rho\sigma}
             +\sfrac 12\psi^\lambda\psi^\nu\5\psi^\sigma\5\psi^\rho
             \6^\mu R_{\nu\lambda\rho\sigma}]
\label{Sigma13}
\eea
where indices $\mu$ of $\Omega$, $R$, $\6$ 
have been raised with the inverse of
$G_{\mu\nu}(x,y)$, and $\psi^i$, $\5\psi^i$ and $\hat F^i$
belong to the same supersymmetry multiplet as $y^i$ (the
auxiliary fields $\hat F^i$
should not be confused with the supercovariant field strengths $\cF^i$
of the gauge fields).
Note that the right hand sides of (\ref{Sigma6}),
(\ref{Sigma7}), (\ref{Sigma8}),
(\ref{Sigma12}) and (\ref{Sigma13})
still contain $\hat F^\mu$, $\5\cD\psi^\mu$ or $\cD\5\psi^\mu$, 
which are to be substituted for by the expressions given in
(\ref{Sigma9}), (\ref{Sigma10}) and (\ref{Sigma11}), 
respectively. Furthermore, in (\ref{Sigma12})
one has to substitute the
expression resulting from (\ref{Sigma13}) for $\cD\5\cD x^\mu$.
Using Eqs.\ (\ref{Sigma1}) through (\ref{Sigma13})
and their $\cD$ and $\5\cD$ derivatives,
we eliminate all variables on the left hand sides of these equations
and all the covariant derivatives of these variables.
Furthermore, we use these equations to define
the $\sigma$-transformations of the remaining field variables
from their $s$-transformations.
For instance, one gets
\bea
\sigma y^i&=& 0
\label{sigmay}\\
\sigma x^\mu &=& (\eta\cD+\5\eta\5\cD)x^\mu +\ep \psi^\mu
                 +\5\ep\5\psi^\mu
\label{sigmax}\\
\sigma\psi^\mu &=& \eta\cD\psi^\mu
             -\sfrac 12\5\eta[\5\cD x^\nu\psi^\rho{\Omega_{\rho\nu}}^\mu
             +\sfrac 12 \psi^\lambda\5\psi^\sigma\5\psi^\rho
             {\Omega_{\lambda\sigma}}^\nu{\Omega^\mu}_{\rho\nu}
             +\psi^\nu\5\psi^\rho\5\psi^\sigma
             {R^\mu}_{\nu\sigma\rho}]
\nonumber\\
&&
            +\sfrac 12\6\eta \psi^\mu
            +\ep\cD x^\mu+\sfrac 12\5\ep \psi^\rho\5\psi^\nu
            {\Omega_{\rho\nu}}^\mu.
\label{sigmapsi}
\eea
The $\sigma$-transformations of $\eta$, $\5\eta$, $\ep$,
$\5\ep$, $\mu$, $\5\mu$, $\alpha$, $\5\alpha$
coincide with their $s$-transformations. The cohomology
$H(\sigma)$ is the cohomology of $\sigma$ in the space of local functions
of the variables $\{u^\ell,v^\ell,W^A\}$, where the $u$'s
and $v$'s are the same as in sections \ref{covfields}
and \ref{action}, while the $W$'s are given by
\bea
\{W^A\}=\{
y^i,x^\mu,\cD^k x^\mu,\5\cD^k x^\mu,
\cD^r \psi^\mu,\5\cD^r \5\psi^\mu,\6^r\eta,\5\6^r\5\eta,
\6^r\ep,\5\6^r\5\ep,C^i:
\nonumber\\
k=1,2,\dots\ ,\ r=0,1,\dots\}.
\label{vars}
\eea
$H(\sigma)$ is well-defined because $\sigma$ squares to zero,
\begin{equation}
\sigma^2=0.
\label{sigma^2}
\end{equation} 
This holds because the (covariant)
equations of motion of the fields
$X$, $\psi$, $\5\psi$, $\hat F$, $\hat\phi$, $\lambda$, 
$\5\lambda$, $A_m$ and their covariant derivatives 
transform into each other 
under diffeomorphisms and supersymmetry transformations
but not into the equations of motion of $\mu$, $\5\mu$, $\alpha$ or
$\5\alpha$ [as can be read off from the $s$-transformations of the
superconformal antifields in appendix \ref{transformations2}]. 

\subsection{Relation to $H(\sigma,\cW)$}

$\sigma$ acts on the variables $\{u^\ell,v^\ell,W^A\}$ 
according to $\sigma u^\ell=v^\ell$, $\sigma W^A=r^A(W)$. 
Furthermore, analogously to (\ref{CONTRACT}) one has
\begin{equation}
\Big\{\sigma\, ,\, \frac{\6}{\6(\6\eta)}\,\Big\}W^A
=L_0 W^A\ ,\ 
\Big\{\sigma\, ,\, \frac{\6}{\6(\5\6\5\eta)}\,\Big\}W^A
=\5L_0 W^A \ ,
\label{H02}
\end{equation}
i.e., in the space of local functions of the $W$'s
the derivatives with respect to $\6\eta$ and $\5\6\5\eta$
are contracting homotopies for $L_0$ and $\5L_0$, respectively.
Hence, the same standard arguments, which were used already
in section \ref{action} yield that
$H(\sigma)$ is given by 
$H_\mathrm{dR}(GL^+(2))\otimes H(\sigma,\cW)$, where
$H_\mathrm{dR}(GL^+(2))$ reflects the nontrivial
de Rham cohomology of the zweibein manifold
(see theorem 5.1 of \cite{Barnich:1995ap}), while
$H(\sigma,\cW)$ is the $\sigma$-cohomology 
in the space
of local functions with vanishing conformal weights made solely
of the variables (\ref{vars}),
\begin{equation}
H(\sigma)=H_\mathrm{dR}(GL^+(2))\otimes H(\sigma,\cW),\quad
\cW=\{\omega:\omega=\omega(W),\ (L_0\omega,\5L_0\omega)=(0,0)\}.
\label{HW}
\end{equation}
The factor $H_\mathrm{dR}(GL^+(2))$ is irrelevant for the
following discussion because it just reflects $\det e_m^a\neq 0$
and makes no difference between superstring and bosonic string
models.

\subsection{Decomposition of $\sigma$}

To study $H(\sigma,\cW)$ we decompose $\sigma$ into pieces
of definite degree in the supersymmetry ghosts
and the fermions\footnote{We are referring here to the
variables (\ref{vars}) themselves, and not to the fermions that are
implicitly contained in these variables through covariant derivatives.}.
The corresponding counting operator is denoted by $N$,
\begin{equation}
N=N_{\ep}+N_{\5\ep}+N_{\psi}+N_{\5\psi}
\label{N}
\end{equation}
with $N_{\ep}$ and $N_{\5\ep}$ as in (\ref{Nep}) and
\[
N_{\psi}=\sum_{r\geq 0}(\cD^r\psi^\mu)\,\frac{\6}{\6(\cD^r\psi^\mu)}\
,\quad
N_{\5\psi}=\sum_{r\geq  0}
(\5\cD^r\5\psi^\mu)\,\frac{\6}{\6(\5\cD^r\5\psi^\mu)}\ .
\]
Using the formulae given above, it is easy to verify that
$\sigma$ decomposes into pieces with even $N$-degree,
\begin{equation}
\sigma=\sum_{n\geq 0}\sigma_{2n}\quad,\quad
[N,\sigma_{2n}]=2n\,\sigma_{2n}
\label{sigmasplit}
\end{equation}
where, on each variable (\ref{vars}),
only finitely many $\sigma_{2n}$ are non-vanishing.
For instance, (\ref{sigmapsi}) yields
\beann
\sigma_0\psi^\mu&=&
\eta\cD\psi^\mu-\sfrac 12\5\eta\5\cD x^\nu\psi^\rho{\Omega_{\rho\nu}}^\mu
+\sfrac 12\6\eta \psi^\mu+\ep\cD x^\mu
\\
\sigma_2\psi^\mu&=&
-\sfrac 14\5\eta\psi^\lambda\5\psi^\sigma\5\psi^\rho
             {\Omega_{\lambda\sigma}}^\nu{\Omega^\mu}_{\rho\nu}
             -\sfrac 12\5\eta\psi^\nu\5\psi^\rho\5\psi^\sigma
             {R^\mu}_{\nu\sigma\rho}
             +\sfrac 12\5\ep \psi^\rho\5\psi^\nu
             {\Omega_{\rho\nu}}^\mu
\\
\sigma_{2n}\psi^\mu&=&0\quad\mbox{for}\quad n>1.
\eeann

\subsection{Decomposition of $\sigma_0$}

We shall prove the asserted result by an inspection of
the cohomology of $\sigma_0$. To that end we decompose
$\sigma_0$ according to the supersymmetry ghosts.
That decomposition has only two pieces owing to the very definition
of $\sigma_0$ and $N$,
\begin{equation}
\sigma_0=\sigma_{0,0}+\sigma_{0,1}\ ,\quad
[N_{\ep}+N_{\5\ep}\,,\sigma_{0,0}]=0\ ,\quad
[N_{\ep}+N_{\5\ep}\,,\sigma_{0,1}]=\sigma_{0,1}\ .
\label{sigma0}
\end{equation}
One easily verifies by induction that $\sigma_{0,1}$ has the following
simple structure:
\bea
\sigma_{0,1}y^i&=&0
\nonumber\\
\sigma_{0,1}\cD^rx^\mu&=&0
\nonumber\\
\sigma_{0,1}\5\cD^rx^\mu&=&0
\nonumber\\
\sigma_{0,1}\cD^r\psi^\mu&=&
\sum_{k=0}^r{r\choose k}\6^k\ep\, \cD^{r+1-k}x^\mu
\nonumber\\
\sigma_{0,1}\5\cD^r\5\psi^\mu&=&
\sum_{k=0}^r{r\choose k}\5\6^k\5\ep\, \5\cD^{r+1-k}x^\mu
\nonumber\\
\sigma_{0,1}\6^r\eta&=&0
\nonumber\\
\sigma_{0,1}\5\6^r\5\eta&=&0
\nonumber\\
\sigma_{0,1}\6^r\ep&=&0
\nonumber\\
\sigma_{0,1}\5\6^r\5\ep&=&0
\nonumber\\
\sigma_{0,1}C^i&=&0.
\label{sigma01}
\eea

\subsection{$H(\sigma_0,\cW)$ at ghost numbers $<5$}

The cocycle condition of $H(\sigma_0,\cW)$ reads
\begin{equation}
\sigma_0\omega=0,\quad \omega\in\cW.
\label{H01}
\end{equation}
We analyse (\ref{H01}) using
(\ref{sigma0}). To that end we decompose $\omega$ according to
the number of supersymmetry ghosts,
\begin{equation}
\omega=\sum_{k=\uk}^{\ok}\omega_k\ ,\quad 
(N_{\ep}+N_{\5\ep})\omega_k=k\omega_k\ .
\label{H04}
\end{equation}
Note that $\ok$ is finite, $\ok\leq \gh(\omega)$. Hence,
the cocycle condition (\ref{H01}) decomposes into
\begin{equation}
\sigma_{0,1}\omega_{\ok}=0,\quad 
\sigma_{0,0}\omega_{\ok}+\sigma_{0,1}\omega_{\ok-1}=0, \quad 
\dots\quad ,\quad \sigma_{0,0}\omega_{\uk}=0.
\label{H05}
\end{equation}
We can neglect contributions $\sigma_{0,1}\7\omega_{\ok-1}$ 
to $\omega_{\ok}$ because such contributions can be removed
by subtracting $\sigma_0\7\omega_{\ok-1}$ from $\omega$.
Hence, $\omega_{\ok}$ can be assumed to be a nontrivial representative
of $H(\sigma_{0,1},\cW)$. That cohomology is computed
in appendix \ref{Hs01} and yields
\bea
\omega_{\ok}
&=&
h(y,x,C,[\ep,\eta],[\5\ep,\5\eta])
+\eta\cD x^\mu h_\mu (y,x,\6\eta,C,[\5\ep,\5\eta])
\nonumber\\
&&
+\5\eta\5\cD x^\mu \5h_\mu (y,x,\5\6\5\eta,C,[\ep,\eta])
+\eta\5\eta\cD x^\mu\5\cD x^\nu
h_{\mu\nu}(y,x,\6\eta,\5\6\5\eta,C)
\label{H20}
\eea
where $\sigma_{0,1}$-exact pieces have been neglected, and
$[\ep,\eta]$ and $[\5\ep,\5\eta]$ denote dependence
on the variables $\6^r\ep,\6^r\eta$ and $\5\6^r\5\ep,\5\6^r\5\eta$
($r=0,1,\dots$), respectively.
The result (\ref{H20}) holds for all ghost numbers
and shows in particular that $\omega_{\ok}$ can be assumed not
to depend on the fermions ($\cD^r\psi^\mu$, $\5\cD^r\5\psi^\mu$)
at all.
We now insert this result in the second equation (\ref{H05}), 
which requires that $\sigma_{0,0}\omega_{\ok}$ be $\sigma_{0,1}$-exact.
At ghost numbers $<5$ this requirement kills completely the dependence
of $\omega_{\ok}$ on the supersymmetry ghosts as we show
in appendix \ref{cases}. The result for these ghost numbers
is thus that, modulo $\sigma_0$-exact pieces, the solutions
to (\ref{H01}) neither depend on the fermions nor on the
supersymmetry ghosts,
\bea
&&\gh(\omega)<5:\ 
\omega=\sigma_0\7\omega+h(y,x,C,[\eta],[\5\eta])
+\eta\cD x^\mu h_\mu (y,x,\6\eta,C,[\5\eta])
\nonumber\\
&&\phantom{\gh(\omega)<5:\ \omega=}
+\5\eta\5\cD x^\mu \5h_\mu (y,x,\5\6\5\eta,C,[\eta])
+\eta\5\eta\cD x^\mu\5\cD x^\nu
h_{\mu\nu}(y,x,\6\eta,\5\6\5\eta,C).\quad\quad\quad
\label{H06}
\eea
Furthermore, (\ref{sigma0}) and (\ref{sigma01}) show that
a function which neither depends
on the fermions nor on the supersymmetry ghosts is
$\sigma_0$-exact if and only if it is the $\sigma_{0}$-transformation
of a function which does not depend on these variables either.
Combining this with (\ref{H06}) one concludes
\begin{equation}
g<5:\quad H^g(\sigma_0,\cW)\simeq H^g(\sigma_{0},\cW_0),
\label{isosigma}
\end{equation}
where $\cW_0$ is the subspace of $\cW$ containing the functions
with vanishing $N$-eigenvalues,
\[
\cW_0=\{\omega\in\cW:N\omega=0\}.
\]
This subspace can be made very explicit. The only variables
(\ref{vars}) with negative conformal weights on which
a function $\omega\in\cW_0$ can depend
are the undifferentiated ghosts $\eta$ and $\5\eta$ [note:
the only other variables (\ref{vars}) with negative conformal weights
are the undifferentiated supersymmetry ghosts, but they do not occur
in $\omega\in\cW_0$ by the very definition of $\cW_0$].
Since $\eta$ and $\5\eta$ are anticommuting variables each of them
can occur at most once in a monomial contributing to 
$\omega\in\cW_0$. Hence, since $\eta$ and $\5\eta$ have
conformal weights $(-1,0)$ and $(0,-1)$, respectively, 
functions in $\cW_0$ can 
only depend on those $w$'s with conformal weights $\leq 1$
(as higher weights cannot be compensated for by variables with
negative weights), and a
variable with $L_0$-weight ($\5L_0$-weight) 1 must necessarily occur
together with $\eta$ ($\5\eta$). This yields
\begin{equation}
\omega\in \cW_0\ \Leftrightarrow\ 
\omega=f(y,x,C,\6\eta,\5\6\5\eta,
\eta\cD x^\mu,\5\eta\5\cD x^\mu,
\eta\6^2\eta,\5\eta\5\6^2\5\eta).
\label{W}
\end{equation}
Note that $H(\sigma_{0},\cW_0)$ is nothing but the
on-shell cohomology $H(\sigma,\cW)$ of the corresponding
bosonic string model, since elements of $\cW_0$ neither depend
on the fermions nor on the supersymmetry ghosts, and since
$\sigma_{0}$ reduces in $\cW_0$ to $\sigma_{0,0}$, which
encodes only the diffeomorphism transformations but not the
supersymmetry transformations.

\subsection{$H(\sigma)$ at ghost numbers $<4$}

We shall now show that $H(\sigma,\cW)$ is at ghost numbers $<4$
isomorphic to $H(\sigma_0,\cW_0)$, 
\begin{equation} 
g<4:\quad H^g(\sigma,\cW)\simeq H^g(\sigma_0,\cW_0).
\label{HsigmaA}
\end{equation}
Because of (\ref{HW})
this implies that $H(\sigma)$ is isomorphic to its
counterpart in the corresponding bosonic string model
(recall that the factor $H_\mathrm{dR}(GL^+(2))$ 
is present in the case of
bosonic strings as well, and that $H^g(\sigma_0,\cW_0)$
is the on-shell cohomology of the bosonic string model).
To derive (\ref{HsigmaA}), we consider
the cocycle condition of $H(\sigma,\cW)$,
\begin{equation}
\sigma\omega=0,\quad \omega\in\cW.
\label{H07}
\end{equation}
We decompose $\omega$ into pieces with definite degree
in the supersymmetry ghosts and fermions,
\begin{equation}
\omega=\sum_{n=\UN}^\ON\omega_n,\quad N\omega_n=n\,\omega_n,
\label{H08}
\end{equation}
with $N$ as in (\ref{N}) [actually there are only even values of $n$ 
in this decomposition because
$\omega$ has vanishing conformal weights]. 
The cocycle condition
(\ref{H07}) implies in particular
\begin{equation}
\sigma_0\omega_\UN=0,
\label{H09}
\end{equation}
where we used the decomposition
(\ref{sigmasplit}) of $\sigma$. Hence, every cocycle
$\omega$ of $H^g(\sigma,\cW)$ contains a coycle
$\omega_\UN$ of $H^g(\sigma_0,\cW)$. Our 
result (\ref{isosigma}) on $H^g(\sigma_0,\cW)$ 
implies that this relation between representatives 
of $H^g(\sigma,\cW)$ and $H^g(\sigma_0,\cW)$ gives rise to
a one-to-one correspondence between the cohomology classes
of  $H^g(\sigma,\cW)$ and $H^g(\sigma_0,\cW_0)$ for $g<4$
and thus to (\ref{HsigmaA}). The arguments are standard
and essentially the following:

(i) When $g<5$,
$\omega_\UN$ can be assumed to be nontrivial
in $H^g(\sigma_0,\cW)$ and represents thus a class of 
$H^g(\sigma_0,\cW_0)$. Indeed, assume it were trivial, i.e.,
$\omega_\UN=\sigma_0\7\omega_\UN$ for some $\7\omega_\UN\in\cW$.
In that case we can
remove $\omega_\UN$ from $\omega$ by subtracting $\sigma\7\omega_\UN$.
$\omega':=\omega-\sigma\7\omega_\UN\in\cW$ is cohomologically equivalent
to $\omega$ and its decomposition (\ref{H08}) starts at some
degree $\UN'>\UN$ unless it vanishes
(which implies already $\omega=\sigma\7\omega_\UN$). 
The cocycle condition for $\omega'$ 
implies $\sigma_0\omega'_{\UN'}=0$ and thus $\omega'_{\UN'}=\sigma_0
\7\omega'_{\UN'}$ for some $\7\omega'_{\UN'}\in\cW$ as a consequence of
(\ref{isosigma}) (owing to $\UN'>\UN\geq 0$).
Repeating the arguments, one concludes that $\omega$ is
$\sigma$-exact, $\omega=\sigma(\7\omega_\UN+\7\omega'_{\UN'}+\dots)$
[it is guaranteed that the procedure terminates, i.e., that the sum
$\7\omega_\UN+\7\omega'_{\UN'}+\dots$ is finite and thus local, because
the number of supersymmetry ghosts is bounded by the ghost number 
and thus the number of fermions is bounded too because
$\omega$ has vanishing conformal weights]. 

(ii) 
When $g<4$, every nontrivial cocycle $\omega_0$ of
$H^g(\sigma_0,\cW_0)$ can be completed to a nontrivial cocycle
$\omega$ of $H^g(\sigma,\cW)$.
Indeed suppose we had constructed
$\omega_n\in\cW$, $n=0,\dots,m$ with ghost number $g$ such that
$\omega^{(m)}:=\sum_{n=0}^m\omega_n$ fulfills
$\sigma\omega^{(m)}=\sum_{n\geq m+1}R_{n}$ with $NR_n=n R_n$
[for $m=0$ this is implied by $\sigma_0\omega_0=0$ which holds
because $\omega_0$ is a $\sigma_0$-cocycle by assumption].
$\sigma^2=0$ implies $\sigma\sum_{n\geq m+1}R_{n}=0$ and thus
$\sigma_0R_{m+1}=0$ at lowest $N$-degree.
Note that $R_{m+1}$ is in $\cW$ (owing to $\sigma\cW\subset\cW$)
and that it has ghost number $g+1<5$ because $\omega^{(m)}$ has
ghost number $g<4$. (\ref{isosigma}) guarantees thus that there is some
$\omega_{m+1}\in\cW$ such that $R_{m+1}=-\sigma_0\omega_{m+1}$, 
which implies that $\omega^{(m+1)}:=\omega^{(m)}+\omega_{m+1}$
fulfills $\sigma\omega^{(m+1)}=\sum_{n\geq m+2}R'_n$. By induction
this implies that every solution to (\ref{H09}) with ghost number $<4$ 
can indeed be completed to a solution of (\ref{H07}) [the locality
of $\omega$ holds by the same arguments as above].
If $\omega_0$ is trivial in $H^g(\sigma_0,\cW_0)$, then
its completion $\omega$ is trivial in $H^g(\sigma,\cW)$ by
arguments used in (i). Conversely, the triviality
of $\omega$ in $H^g(\sigma,\cW)$ ($\omega=\sigma\eta$) implies
obviously the triviality of $\omega_0$ 
in $H^g(\sigma_0,\cW_0)$ ($\omega_0=\sigma_0\eta_0$)
because there are no negative $N$-degrees.

\mysection{Relation to the cohomology of bosonic strings}\label{ISO}

We shall now derive (\ref{iso}) and the
announced isomorphism between the $s$-cohomologies
of a superstring and the corresponding bosonic
string model. Both results can be traced
to the existence of variables $\{\4u^{\4\ell},\4v^{\4\ell},\4W^{\4A}\}$
on which $s$ takes a form very similar to $\sigma$
on the variables $\{u^{\ell},v^{\ell},w^A\}$ 
used in section \ref{sigmacoho}. In the `Beltrami basis'
the set of $\4u$'s consists of: (i) $\4u$'s with
ghost number 0 which coincide with the $u^\ell$; 
(ii) $\4u$'s with
ghost number $-1$ given by the 
superconformal antifields $X^*_M$, $\psi^*_M$,
$\5\psi^*_M$, $F^*_M$, $\phi^*_i$, $\lambda^*_i$
$\5\lambda^*_i$, $A^*_i$ (recall that we have dropped
the hats on these antifields) and all covariant
derivatives of these antifields plus
the $\5A^*_i$ and all their $\5\cD$-derivatives 
($\5\cD^r\5A^*_i$, $r=0,1,\dots$)\footnote{The $\cD^k\5\cD^r\5A^*_i$
with $k>0$ do not count among the $u$'s because the antifield
independent parts of $s\cD^k\5\cD^r\5A^*_i$
and $-s\cD^{k-1}\5\cD^{r+1}A^*_i$ are equal
(both are given by $\cD^k\5\cD^{r+1}y^i$).
Rather, they are substituted for by the $v$'s corresponding to
the $\cD^{k-1}\5\cD^{r}C^*_i$ ($k>0$) owing to 
$s\cD^{k-1}\5\cD^{r}C^*_i=-\cD^k\5\cD^r\5A^*_i+\dots$.};
(iii) $\4u$'s with ghost number $-2$ given by the antifields
of the ghosts, i.e., by
$\eta^*$, $\5\eta^*$,
$\ep^*$, $\5\ep^*$, $C^*_i$ and all their
derivatives. It can be readily checked that
a complete set of new local jet coordinates in the Beltrami basis
is given by $\{\4u^{\4\ell},\4v^{\4\ell},\4W_{(0)}^{\4A}\}$
with $\4v^{\4\ell}=s\4u^{\4\ell}$ and 
\bea
\{\4W_{(0)}^{\4A}\}=\{
y^i,x^\mu,\cD^k x^\mu,\5\cD^k x^\mu,
\cD^r \psi^\mu,\5\cD^r \5\psi^\mu,\6^r\eta,\5\6^r\5\eta,
\6^r\ep,\5\6^r\5\ep,C^i,
\nonumber\\
\6^r\mu^*,\5\6^r\5\mu^*,\6^r\alpha^*,\5\6^r\5\alpha^*:
k=1,2,\dots\ ,\ r=0,1,\dots\}.
\label{w0}
\eea
Note that $\{\4W_{(0)}^{\4A}\}$ does not only contain
the $W^A$ listed in (\ref{vars}), but in addition the variables $\6^r\mu^*$, 
$\5\6^r\5\mu^*$, $\6^r\alpha^*$, $\5\6^r\5\alpha^*$.
The latter occur here because
their $s$-transformations contain no linear parts and can therefore
not be used as $\4v$'s\footnote{The other
derivatives of the antifields
$\mu^*$, $\5\mu^*$, $\alpha^*$, $\5\alpha^*$, such as the
$\5\6^k\6^r\mu^*$ ($k>0$), do not occur among the $\4W_{(0)}$'s
because they are substituted for by the $\4v$'s corresponding
to $\eta^*$, $\5\eta^*$, $\ep^*$, $\5\ep^*$ and their
derivatives (e.g., one has $s\eta^*=-\5\6\mu^*+\dots$).}.
The $\4W_{(0)}^{\4A}$ fulfull
\begin{equation}
s\4W_{(0)}^{\4A}=r^{\4A}(\4W_{(0)})+O(1)
\label{sw0}
\end{equation}
where $O(1)$ collects terms which are at least linear in the
$\4u$'s and $\4v$'s. As shown in \cite{Brandt:2001tg}, 
(\ref{sw0}) implies the existence of
variables $\4W^{\4A}=\4W_{(0)}^{\4A}+O(1)$ which fulfill
\begin{equation}
s\4W^{\4A}=r^{\4A}(\4W)
\label{w}
\end{equation}
with the {\em same} functions $r^{\4A}$ as in (\ref{sw0}).
Furthermore the algorithm described
in \cite{Brandt:2001tg} for the construction of the
$\4W^{\4A}$ results in local expressions when applied
in the present case.
This can be shown by means of arguments similar to those
used within the discussion of the examples
in \cite{Brandt:2001tg}%
\footnote{In the present case, the suitable `degrees' to be
used in these arguments are the conformal 
weights and the ghost number. Using these degrees
one can prove that the algorithm produces local (though
not necessarily polynomial)
expressions: the resulting
$\4W$'s can depend nonpolynomially on the $x^\mu$, $y^i$
and on the two particular combinations $\ep\5\lambda^*_i$ and
$\5\ep\lambda^*_i$ but they are necessarily
polynomials in all variables which contain derivatives
of fields or antifields.}.

(\ref{w}) implies that the 
$s$-transformations of those $\4W$'s which correspond to
the variables (\ref{vars}) can be obtained from the
$\sigma$-transformations of the latter variables simply
by substituting there $\4W$'s for the corresponding $W$'s. 
For instance, this gives
\bea
sy^{i\prime}&=&0,
\label{sy'}\\
sx^{\mu\prime} &=& \eta(\cD x^\mu)^\prime+
\5\eta(\5\cD x^\mu)^\prime+\ep \psi^{\mu\prime}+\5\ep \5\psi^{\mu\prime}
\label{sx'}
\eea
where here and in the following a prime on a variable indicates a 
$\4W$-variable\footnote{The construction of the $\4W$'s implies
$(\6^r\eta)'=\6^r\eta$, $(\5\6^r\5\eta)'=\5\6^r\5\eta$,
$(\6^r\ep)'=\6^r\ep$ and $(\5\6^r\5\ep)'=\5\6^r\5\ep$
because the $s$-transformation of these ghost variables do not
contain any $\4u$'s or $\4v$'s. This has been used in (\ref{sx'}).}.
For instance, $y^{i\prime}$ is the $\4W$-variable
corresponding to $y^i$ and explicitly given by
\begin{equation}
y^{i\prime}=y^i+\ep\5\lambda^*_i-\5\ep\lambda^*_i
-\eta A^*_i+\5\eta \5A^*_i +\eta\5\eta C^*_i\ .
\label{y'}
\end{equation}
This very close relation between $s$ on the $\4W$-variables and
$\sigma$ on the variables (\ref{vars}) would immediately imply
$H(s)\simeq H(\sigma)$ if the $\4W$-variables
$(\6^r\mu^*)'$, $(\5\6^r\5\mu^*)'$, $(\6^r\alpha^*)'$,
$(\5\6^r\5\alpha^*)'$ were not present. 
Nevertheless the asserted isomorphism 
(\ref{iso}) holds because the conformal
weights of the latter variables are too high so that they
cannot contribute nontrivially to 
$H^g(s)$ for $g<4$.
To show this we analyse $H(s)$
along the same lines as $H(\sigma)$ in section
\ref{sigmacoho}. 

The first step of that analysis gives 
\begin{equation}
H(s)\simeq H_\mathrm{dR}(GL^+(2))\otimes
H(s,\4\cW),\quad \4\cW=\{\omega:\omega=\omega(w),\
(L_0\omega,\5L_0\omega)=(0,0)\}.
\label{Arschloch}
\end{equation}
This result is analogous to (\ref{HW}) and expresses that
the zweibein gives the only nontrivial cohomology in the
subspace of $\4u$'s and $\4v$'s and that
there is a contracting homotopy for $L_0$ and $\5L_0$ because
(\ref{w}) implies
\[
\Big\{s\, ,\, \frac{\6}{\6(\6\eta)}\,\Big\} \4W^{\4A}=L_0 \4W^{\4A}\ ,\ 
\Big\{s\, ,\, \frac{\6}{\6(\5\6\5\eta)}\,\Big\}\4W^{\4A}=\5L_0 \4W^{\4A}\ .
\]
The conformal weights of $\alpha^{*\prime}$,
$\5\alpha^{*\prime}$, $\mu^{*\prime}$ and $\5\mu^{*\prime}$
are $(3/2,0)$, $(0,3/2)$, $(2,0)$ and $(0,2)$, respectively.

$H(s,\4\cW)$ can be analysed by means of a decomposition of $s$
analogous to the $\sigma$-decomposition in (\ref{sigmasplit}),
using a counting operator
$N'$ for all those $\4W$'s which have half-integer conformal weights,
\[
N'=N_\ep+N_{\5\ep}+N_{\psi'}+N_{\5\psi'}+N_{\alpha^{*\prime}}
+N_{\5\alpha^{*\prime}}\ .
\]
The decomposition of $s$ reads
\[
s=\sum_{n\geq 0}s_{2n}\quad,\quad
[N',s_{2n}]=2n\,s_{2n}\ .
\]
Next we examine the $s_0$-cohomology. 
Analogously to (\ref{sigma0}) one has
\[
s_0=s_{0,0}+s_{0,1}\ ,\quad
[N_{\ep}+N_{\5\ep}\,,s_{0,0}]=0\ ,\quad
[N_{\ep}+N_{\5\ep}\,,s_{0,1}]=s_{0,1}\ .
\]
We now determine the cohomology of $s_{0,1}$ along the lines
of the investigation of the $\sigma_{0,1}$-cohomology in appendix
\ref{Hs01} by inspecting the part of $s_{0,1}$
which contains the undifferentiated ghost $\ep$. That part is the
analog of $\sigma_{0,1,1}$ in (\ref{H088}) and
takes the form $\ep\, \7G'_{-1/2}$.
$\7G'_{-1/2}$ acts nontrivially only on the
$\psi'$, $\alpha^{*\prime}$ and their (covariant) derivatives
according to
\[
\7G'_{-1/2}(\cD^r\psi^\mu)'= (\cD^{r+1}x^\mu)'\quad ,\quad
\7G'_{-1/2}(\6^r\alpha^*)'= -(\6^r\mu^*)'
\]
We define
a contracting homotopy $B'$ which is analogous to the
contracting homotopy $B$ in appendix \ref{Hs01}, 
\[
B'=\sum_{r\geq 0}\Big[(\cD^{r}\psi^\mu)'\, 
\frac{\6}{\6(\cD^{r+1}x^\mu)'}-
(\6^r\alpha^*)'\,\frac{\6}{\6(\6^r\mu^*)'}
\Big].
\]
Using
$B'$ one proves that the functions $f'_m$ with $m>0$ which are analogous
to the functions $f_m$ in appendix \ref{Hs01} can be assumed not to depend
on the variables $(\cD^r\psi^\mu)'$, $(\cD^{r+1}x^\mu)'$, 
$(\6^r\alpha^*)'$ or $(\6^r\mu^*)'$.%
\footnote{For this argument it is important that
there is a finite maximal value $\OR$ of $m$. In the case of
the $\sigma$-cohomology,
$m$ was bounded from above by the ghost number but now the ghost number
alone does not give a bound because there are variables with negative
ghost numbers,
the $(\6^r\alpha^*)'$, $(\5\6^r\5\alpha^*)'$, $(\6^r\mu^*)'$ and
$(\5\6^r\5\mu^*)'$.
Nevertheless there is a bound because 
$\omega(\4W)$ does not only have fixed ghost number but also vanishing
conformal weights. Indeed, it is easy to show that this forbids
arbitrarily large powers of $\ep$ because
the $(\6^r\alpha^*)'$ and $(\6^r\mu^*)'$
have ghost number $-1$ and conformal weights $\geq 3/2$.} 
In the case $m=0$ one 
gets that $f'_0$ does not depend
on $(\6^r\alpha^*)'$ or $(\6^r\mu^*)'$, 
simply because the conformal weights of
these variables are too large [cf.\ the
arguments in the text after (\ref{H15})]. 
This implies the analog of equation (\ref{H17}),
with functions $h'_m$ and $g'_\mu$ which may still depend on
$(\5\cD^r\5\psi^\mu)'$, $(\5\cD^{r+1}x^\mu)'$, 
$(\5\6^r\5\alpha^*)'$ or $(\5\6^r\5\mu^*)'$.
The dependence on these variables can be analysed 
analogously,
using a contracting homotopy $\5B'$ for these variables,
along the lines of the remaining analysis in appendix \ref{Hs01}.
One finally obtains the following result for $H(s_{0,1},\4\cW)$:
\bea
&&
s_{0,1}\omega=0,\quad \omega\in\4\cW\quad \then\ 
\nonumber\\
&&
\omega=
h(y',x',C',[\ep,\eta],[\5\ep,\5\eta])
\nonumber\\
&&
\phantom{\omega=}
+\eta(\cD x^\mu)^{\prime} h_\mu (y',x',\6\eta,C',[\5\ep,\5\eta])
+\5\eta(\5\cD x^\mu)^{\prime} \5h_\mu (y',x',\5\6\5\eta,C',[\ep,\eta])
\nonumber\\
&&
\phantom{\omega=}
+\eta\5\eta(\cD x^\mu)^{\prime}(\5\cD x^\nu)^{\prime}
h_{\mu\nu}(y',x',\6\eta,\5\6\5\eta,C')
+s_{0,1}\7\omega(w),\ \7\omega\in\4\cW.
\label{wichtig}
\eea
Hence, $H(s_{0,1},\4\cW)$ is completely isomorphic to
$H(\sigma_{0,1},\cW)$
(for all ghost numbers).
In particular, the representatives do not depend
on $(\6^r\alpha^*)'$, $(\5\6^r\5\alpha^*)'$,
$(\6^r\mu^*)'$ or $(\5\6^r\5\mu^*)'$ [recall that the reason is
that the conformal weights of these variables are too high; if,
for instance, $\mu^{*\prime}$ had conformal weights $(1,0)$ 
instead of $(2,0)$
it had contributed to (\ref{wichtig}) analogously to
$(\cD x^\mu)^{\prime}$]. This implies
the results announced above: arguments which are 
completely analogous to those used to derive first (\ref{H06}) and
then (\ref{HsigmaA}) lead to
\begin{equation}
g<4:\quad H^g(s,\4\cW)\simeq H^g(s_0,\4\cW_0),\quad
\4\cW_0=\{\omega\in\4\cW:N'\omega=0\}.
\label{isos}
\end{equation}
Analogously to (\ref{W}), the elements of $\4\cW_0$
can only depend on those $w$'s with conformal weights $\leq 1$, i.e.,
\begin{equation}
\omega'\in \4\cW_0\ \Leftrightarrow\ 
\omega'=f(y',x',C',\6\eta,\5\6\5\eta,
\eta(\cD x^\mu)^{\prime},
\5\eta(\5\cD x^\mu)^{\prime},
\eta\6^2\eta,\5\eta\5\6^2\5\eta).
\label{W'}
\end{equation}
Because of (\ref{w}),
$s_0$ takes exactly the same form in $\4\cW_0$ as
$\sigma_0$ in $\cW_0$. This implies (for all ghost numbers)
\begin{equation}
H(s_0,\4\cW_0)\simeq H(\sigma_0,\cW_0).
\label{direct}
\end{equation}
Because of (\ref{isos}) and (\ref{HsigmaA}) (as well as
(\ref{Arschloch}) and (\ref{HW})) this yields (\ref{iso}).
(\ref{isos}) establishes also the equivalence between the cohomologies
of the superstring and the corresponding bosonic
 string at ghost numbers $<4$ because $H_\mathrm{dR}(GL^+(2))\otimes
H(s_0,\4\cW_0)$
is nothing but the $s$-cohomology of the bosonic
string.

\section*{Acknowledgements}

A.K. was supported by the \"ONB under grant number 7731 and by the Austrian 
Research Fund FWF under grant number P14639-TPH.

\clearpage

\appendix
\mysection{Cohomology of $\sigma_{0,1}$ in $\cW$}
\label{Hs01}

In this appendix we compute $H(\sigma_{0,1},\cW)$ where $\sigma_{0,1}$
is given in (\ref{sigma01}).
The cocycle condition reads
\begin{equation}
\sigma_{0,1}\omega=0,\quad \omega\in\cW.
\label{H10a}
\end{equation}
We decompose this equation into pieces with definite
degree in the undifferentiated supersymmetry ghosts $\ep$.
$\sigma_{0,1}$ decomposes into two pieces, $\sigma_{0,1,0}$ and 
$\sigma_{0,1,1}$, where $\sigma_{0,1,0}$ does not change the
degree in the undifferentiated $\ep$, whereas $\sigma_{0,1,1}$ increases
this degree by one unit. $\sigma_{0,1,1}$ reads
\begin{equation}
\sigma_{0,1,1}=\ep\,\7G_{-1/2}\ ,\quad
\7G_{-1/2}=\sum_{r\geq 0}(\cD^{r+1}x^\mu)\, 
\frac{\6}{\6(\cD^{r}\psi^\mu)}\ .
\label{H088}
\end{equation}
$\omega$ can be assumed to have fixed ghost number
and is thus a polynomial in the undifferentiated $\ep$,
\begin{equation}
\omega=\sum_{m=\ur}^\OR\ep^m f_m\ ,
\label{H099}
\end{equation}
where $f_m$ can depend on all variables (\ref{vars}) except
for the undifferentiated $\ep$. At highest degree in
the undifferentiated $\ep$, (\ref{H10a}) implies
$\sigma_{0,1,1}(\ep^\OR f_\OR)=0$ and thus
\begin{equation}
\7G_{-1/2} f_\OR=0.
\label{H10}
\end{equation}
We analyse this condition by means of
the contracting homotopy
\[
B=\sum_{r\geq 0}(\cD^{r}\psi^\mu)\, 
\frac{\6}{\6(\cD^{r+1}x^\mu)}\ .
\]
The anticommutator of $B$ and $\7G_{-1/2}$ is the counting
operator for all variables
$\cD^{r}\psi^\mu$ and $\cD^{r+1}x^\mu$ ($r=0,1,\dots$),
\[
\{B,\7G_{-1/2}\}=\sum_{r\geq 0}\Big[(\cD^{r}\psi^\mu)\, 
\frac{\6}{\6(\cD^{r}\psi^\mu)}+(\cD^{r+1}x^\mu)\, 
\frac{\6}{\6(\cD^{r+1}x^\mu)}\Big] .
\]
Hence, (\ref{H10}) implies by standard arguments that
$f_\OR$ is $\7G_{-1/2}$-exact up to a
function that does not depend on the $\cD^{r}\psi^\mu$ or 
$\cD^{r+1}x^\mu$,
\begin{equation}
f_\OR=\7G_{-1/2}\,g_\OR+
h_\OR(y,x,C,[\5\cD x,\5\psi],[\6\ep,\eta],[\5\ep,\5\eta])
\label{H11}
\end{equation}
where $g_\OR$ is a function that can depend on all variables
(\ref{vars}) except for the undifferentiated $\ep$,
$[\5\cD x,\5\psi]$ denotes collectively
the variables $\5\cD^{r+1}x^\mu,\5\cD^{r}\5\psi^\mu$,
and $[\6\ep,\eta]$ and $[\5\ep,\5\eta]$ denote
collectively the variables
$\6^{r+1}\ep,\6^r\eta$ and $\5\6^r\5\ep,\5\6^r\5\eta$,
respectively ($r=0,1,\dots$ in all cases).
We shall first study the case $\OR>0$ [the case $\OR=0$ will be included
automatically below].
(\ref{H11}) implies
\begin{eqnarray}
\OR>0:\quad
\omega&=&\sigma_{0,1}(\ep^{\OR-1} g_\OR )+\ep^{\OR-1}f'_{\OR-1}+
\sum_{m=\ur}^{\OR-2}\ep^m f_m
\nonumber\\
&&+\ep^\OR h_\OR(y,x,C,[\5\cD x,\5\psi],[\6\ep,\eta],[\5\ep,\5\eta])
\label{H12a}
\end{eqnarray}
where
\[
f'_{\OR-1}=f_{\OR-1}-\sigma_{0,1,0}\, g_\OR\ .
\]
The exact piece $\sigma_{0,1}(\ep^{\OR-1} g_\OR)$ on the right hand side
of (\ref{H12a}) will be neglected in the following, i.e., actually
we shall examine $\omega':=
\omega-\sigma_{0,1}(\ep^{\OR-1} g_\OR)$ in the following.
However, for notational convenience, we shall drop the primes
(of $\omega'$ and $f'_{\OR-1}$) and consider now
\begin{equation}
\OR>0:\quad
\omega=
\sum_{m=\ur}^{\OR-1}\ep^m f_m+
\ep^\OR h_\OR(y,x,C,[\5\cD x,\5\psi],[\6\ep,\eta],[\5\ep,\5\eta])
\label{H12}
\end{equation}
We have thus learned that, if $\OR>0$, the piece in
$\omega$ with highest degree in the undifferentiated
$\ep$ can be assumed not to depend on any of the variables
$\cD^{r}\psi^\mu$ or $\cD^{r+1}x^\mu$ ($r=0,1,\dots$).
As a consquence, the $\sigma_{0,1}$-transformation of that
piece does not depend on these variables either and
$\sigma_{0,1}\omega=0$, with $\omega$ as in
(\ref{H12}), implies
\begin{equation}
\7G_{-1/2} f_{\OR-1}=0.
\label{H14}
\end{equation}
We can now analyse (\ref{H14}) in the same way as (\ref{H10})
and repeat the arguments until we reach an equation
\begin{equation}
\7G_{-1/2} f_{0}=0
\label{H15}
\end{equation}
where $f_{0}$ is a function
with conformal weights $(0,0)$ which does not depend on
the undifferentiated $\ep$ [note that $f_m$
has conformal weights $(m/2,0)$ because $\ep^m f_m$ has
conformal weights $(0,0)$; if $\OR$ had been zero, we had
arrived at (\ref{H15}) immediately]. The only way
in which $f_{0}$ can depend nontrivially on
the variables $\cD^{r}\psi^\mu$ or $\cD^{r+1}x^\mu$
($r=0,1,\dots$) is through terms of the form
$\eta\psi^\mu\psi^\nu f_{\mu\nu}(y,x,\6\eta,C,
[\5\cD x,\5\psi],[\5\ep,\5\eta])$, 
$\eta\6\ep \psi^\mu f_{\mu}(y,x,\6\eta,C,
[\5\cD x,\5\psi],[\5\ep,\5\eta])$, or 
$\eta\cD x^\mu g_\mu(y,x,\6\eta,C,
[\5\cD x,\5\psi],[\5\ep,\5\eta])$
[recall that the only variables (\ref{vars}) with negative
$L_0$-weights are the undifferentiated $\eta$ and $\ep$ and
that $\eta$ is an anticommuting variable].
(\ref{H15}) implies
$f_{\mu\nu}(y,x,\6\eta,C,
[\5\cD x,\5\psi],[\5\ep,\5\eta])=0$ and $f_{\mu}(y,x,\6\eta,C,
[\5\cD x,\5\psi],[\5\ep,\5\eta])=0$. 
We conclude
\begin{equation}
f_{0}=
\eta\cD x^\mu g_\mu(y,x,\6\eta,C,
[\5\cD x,\5\psi],[\5\ep,\5\eta])+
h_0(y,x,C,[\5\cD x,\5\psi],[\6\ep,\eta],[\5\ep,\5\eta])
\label{H16}
\end{equation}
We thus get the following intermediate result: without
loss of generality we can assume
\begin{equation}
\omega=\sum_{m}
\ep^m h_m(y,x,C,[\5\cD x,\5\psi],[\6\ep,\eta],[\5\ep,\5\eta])
+\eta\cD x^\mu g_\mu(y,x,\6\eta,C,
[\5\cD x,\5\psi],[\5\ep,\5\eta]).
\label{H17}
\end{equation}
The only part of $\sigma_{0,1}$ which is active
on such an $\omega$ is the part
\[
\7\sigma_{0,1}=
\sum_{r\geq 0}\sum_{k=0}^r{r\choose k}(\5\6^k\5\ep\, \5\cD^{r+1-k}x^\mu)\,
\frac{\6}{\6(\5\cD^r\5\psi^\mu)}\ .
\]
Note that $\7\sigma_{0,1}$ touches only the dependence on the 
variables $\5\cD^r\5\psi^\mu$, $\5\cD^{r+1}x^\mu$ and
$\5\6^r\5\ep$ ($r=0,1,\dots$) and treats all other variables
as contants. Hence, for $\omega$ as in
(\ref{H17}), $\sigma_{0,1}\omega=0$ implies
\bea
&\7\sigma_{0,1} h_m(y,x,C,[\5\cD x,\5\psi],[\6\ep,\eta],[\5\ep,\5\eta])=0 
\quad \forall m,&
\nonumber\\
&\7\sigma_{0,1}g_\mu(y,x,\6\eta,C,
[\5\cD x,\5\psi],[\5\ep,\5\eta])=0.&
\label{H18}
\eea
These equations are decomposed into pieces with definite degree in the
undifferentiated $\5\ep$ and then analysed using the contracting homotopy
\[
\5B=\sum_{r\geq 0}(\5\cD^{r}\5\psi^\mu)\, 
\frac{\6}{\6(\5\cD^{r+1}x^\mu)}\ .
\]
By means of arguments analogous to those that have led to
(\ref{H17}) we conclude that we can assume, without loss
of generality,
\bea
h_m(y,x,C,[\5\cD x,\5\psi],[\6\ep,\eta],[\5\ep,\5\eta])
&=&
\sum_{q}
\5\ep^q h_{m,q}(y,x,C,[\6\ep,\eta],[\5\6\5\ep,\5\eta])
\nonumber\\
&&
+\5\eta\5\cD x^\mu g_{m,\mu}(y,x,\5\6\5\eta,C,
[\6\ep,\eta]),
\nonumber\\[4pt]
g_\mu(y,x,\6\eta,C,
[\5\cD x,\5\psi],[\5\ep,\5\eta])
&=&
\sum_{q}
\5\ep^q h_{\mu,q}(y,x,\6\eta,C,[\5\6\5\ep,\5\eta])
\nonumber\\
&&
+\5\eta\5\cD x^\nu g_{\mu,\nu}(y,x,C,\6\eta,\5\6\5\eta).
\label{H19}
\eea
Since the $h_{m,q}$, $g_{m,\mu}$, $h_{\mu,q}$ and $g_{\mu,\nu}$ do not depend
on the fermions, they are $\sigma_{0,1}$-invariant.
We have thus proved that (\ref{H10a}) implies
\bea
\omega&=&
h(y,x,C,[\ep,\eta],[\5\ep,\5\eta])
\nonumber\\
&&
+\eta\cD x^\mu h_\mu (y,x,\6\eta,C,[\5\ep,\5\eta])
+\5\eta\5\cD x^\mu \5h_\mu (y,x,\5\6\5\eta,C,[\ep,\eta])
\nonumber\\
&&
+\eta\5\eta\cD x^\mu\5\cD x^\nu
h_{\mu\nu}(y,x,\6\eta,\5\6\5\eta,C)+\sigma_{0,1}\7\omega
\label{H19a}
\eea
where the functions on the right hand side ($h$, 
$\eta\cD x^\mu h_\mu$, \dots , $\7\omega$) are elements of $\cW$.
Note also that the sum on the right hand side is direct: no
nonvanishing function $h+\eta\cD x^\mu h_\mu
+\5\eta\5\cD x^\mu \5h_\mu+\eta\5\eta\cD x^\mu\5\cD x^\nu h_{\mu\nu}$
is $\sigma_{0,1}$-exact because the various terms either do not
contain variables $\cD^{r+1}x^\mu$ or $\5\cD^{r+1}x^\mu$ at all,
or they contain $\cD x^\mu$ but no $\ep$, or 
$\5\cD x^\mu$ but no $\5\ep$.
Hence, our result characterizes $H(\sigma_{0,1},\cW)$ completely.

\mysection{Derivation of (\ref{H06})}
\label{cases}

We shall show that (\ref{H20}) implies (\ref{H06}).
The proof is a case-by-case study for $g=0,\dots,4$.
Since $\omega_\ok$ does not depend on the fermions and
has vanishing conformal weights,
it can be assumed to contain only terms with even
$N_\ep$-degree and even $N_{\5\ep}$-degree.
Hence, it does not depend on the supersymmetry ghosts if
$g=0$ or $g=1$ which gives (\ref{H06}) in these cases. 
If $2\leq g\leq 4$ the assertion follows from 
\begin{equation}
\sigma_{0,0}\omega_\ok+\sigma_{0,1}\omega_{\ok-1}=0,
\label{sex}
\end{equation}
which is the second equation in (\ref{H05}).

\underline{$g=2$:} Only $\omega_{\ok=2}$
can depend on the supersymmetry ghosts. One has
\[
\omega_{\ok=2}=\ep\6\ep a(X)+\5\ep\5\6\5\ep \5a(X)
\]
where $a(X)$ and $\5a(X)$ are functions of the undifferentiated
$x^\mu$ and $y^i$.
$\sigma_{0,0}\omega_{\overline{2}}$ contains for instance
$\eta(\6\ep)^2 a(X)$ and $\5\eta(\5\6\5\ep)^2 \5a(X)$ because
$\sigma_{0,0}\ep$ and $\sigma_{0,0}\5\ep$ contain
$\eta\6\ep$ and $\5\eta\5\6\5\ep$, respectively.
If $a\neq 0$ or $\5a\neq 0$,
these terms are not $\sigma_{0,1}$-exact because they
do not contain derivatives of an $x^\mu$. We conclude that
$a=0$ and $\5a=0$ and thus that (\ref{H06}) holds for $g=2$.

\underline{$g=3$:} Again, only $\omega_{\ok=2}$ 
can depend on the supersymmetry ghosts. The terms in $\omega_{\ok=2}$
depending on $\ep$ or its derivatives are
\bea
\eta\ep\6^2\ep a(X)+
\ep\6\ep\6\eta b(X)+
\ep\6\ep\5\6\5\eta c(X)+
\ep\6\ep C^i d_{i}(X)
\nonumber\\
+\5\eta\5\cD x^\mu \ep\6\ep e_\mu(X)+
\eta(\6\ep)^2 f(X)+\6^2\eta\ep^2g(X).
\label{sex1}
\eea
In addition there are analogous terms with $\5\ep$ or its derivatives.
A straightforward
calculation shows that (\ref{sex}) imposes
\begin{equation}
b=0,\quad c=0,\quad d_i=0,\quad
e_\mu=\6_\mu a,\quad f=a,\quad g=-\sfrac 12 a
\label{sex2}
\end{equation}
where $a=a(X)$ is an arbitary function of the $y^i$ and $x^\mu$.
Using (\ref{sex2}) in (\ref{sex1}), the latter becomes
\bea
&[\eta\ep\6^2\ep+
\5\eta\5\cD x^\mu \ep\6\ep \6_\mu+
\eta(\6\ep)^2 -\sfrac 12\6^2\eta\ep^2]a(X)&
\nonumber\\
&=\sigma_{0}[\ep\6\ep a(X)]+\sigma_{0,1}[\eta\6\ep\psi^\mu\6_\mu a(X)].&
\label{sex3}
\eea
This shows that all terms containing
$\ep$ or its derivatives can be removed from $\omega_{\ok=2}$
by the redefinition 
$\omega'=\omega-\sigma_{0}[\ep\6\ep a(X)+\eta\6\ep\psi^\mu\6_\mu a(X)]$.
Similarly one can remove all terms containing
$\5\ep$ or its derivatives. Hence, without loss of generality one
can assume $\omega_{\ok=2}=0$ which implies (\ref{H06}) for $g=3$.

\underline{$g=4$:} Now $\omega_{\ok=4}$ and
$\omega_{2}$
can depend on the supersymmetry ghosts. One has
\[
\omega_{\ok=4}=\ep^3\6^2\ep a(X)+\ep^2(\6\ep)^2 b(X)+
\5\ep^3\5\6^2\5\ep \5a(X)+\5\ep^2(\5\6\5\ep)^2 \5b(X)+
\ep\6\ep\5\ep\5\6\5\ep c(X).
\]
The fact that $\sigma_{0,0}\6^2\ep$
contains $-(1/2)\ep\6^3\eta$ implies $a=0$.
Analogously one concludes $\5a=0$.
The fact that $\sigma_{0,0}\6\ep$ and $\sigma_{0,0}\5\6\5\ep$
contain $\eta\6^2\ep$ and $\5\eta\5\6^2\5\ep$, respectively,
implies $b=0$, $\5b=0$ and $c=0$.

$\omega_{2}$ is of the form 
$P^A(\mbox{ghosts},\cD x^\mu,\5\cD x^\mu)a_A(X)$ where the $P^A$
either depend on $\ep$ and its derivatives, or on 
$\5\ep$ and its derivatives. The complete list of
polynomials $P^A$ depending on $\ep$ and its derivatives is
\beann
&
\eta\6\eta\ep\6^2\ep,
\eta\6\eta(\6\ep)^2,
\6^2\eta\6\eta\ep^2,
\eta\6^2\eta \ep\6\ep,
\eta\6^3\eta\ep^2,
&
\\
&
\5\eta\5\cD x^\mu \eta\ep\6^2\ep,
\5\eta\5\cD x^\mu \eta(\6\ep)^2,
\5\eta\5\cD x^\mu\6\eta\ep\6\ep,
\5\eta\5\cD x^\mu\6^2\eta\ep^2,
&
\\
&
\eta\5\6\5\eta\ep\6^2\ep,
\eta\5\6\5\eta(\6\ep)^2,
\5\eta\5\6^2\5\eta \ep\6\ep,
\6\eta\5\6\5\eta\ep\6\ep,
\6^2\eta\5\6\5\eta\ep^2,
\5\eta\5\cD x^\mu\5\6\5\eta\ep\6\ep,
&
\\
&
\eta C^i\ep\6^2\ep,
\eta C^i(\6\ep)^2,
\6\eta C^i\ep\6\ep,
\6^2\eta C^i\ep^2,
\5\6\5\eta C^i \ep \6\ep,
\5\eta\5\cD x^\mu C^i\ep\6\ep,
C^i C^j\ep\6\ep,
&
\eeann
Starting with the terms
\begin{equation}
\label{two}
\ep\6^2\ep\eta\6\eta A_1(X) + 
(\6\ep)^2\eta\6\eta B_1(X) + 
\ep^2\6\eta\6^2\eta E_2(X)
\end{equation}
one finds that (\ref{sex}) implies $A_1(X)=B_1(X)=2 E_2(X)$. Considering 
the terms 
\begin{eqnarray}
\label{three}
&
\ep\6\ep\eta\6^2\eta B_5(X) + 
\ep^2\eta\6^3\eta E_1(X) + 
\ep\6^2\ep\eta\5\eta\5\cD x^{\mu} A_{4,\mu}(X) \nn
&
\\
& 
+(\6\ep)^2\eta\5\eta\5\cD x^{\mu} B_{4,\mu}(X) + 
\ep\6\ep\6\eta\5\eta\5\cD x^{\mu} C_{4,\mu}(X) + 
\ep^2\6^2\eta\5\eta\5\cD x^{\mu} E_{6,\mu}(X),
&
\end{eqnarray}
one observes that the $\sigma_{0}$ transformation of these terms neither 
contain 
$\5\6^k\5\eta$ or $\5\6^k\5\ep$ terms nor $U(1)$ ghosts. Thus they have to 
fulfill 
(\ref{sex}) separately and one obtains
\begin{eqnarray}
C_{4,\mu}(X)&=&-\6_{\mu} A_1(X) \nn
\\
B_{4,\mu}(X)&=&-\6_{\mu}B_5(X)+\6_{\mu} A_1(X)-2E_{6,\mu}(X) \nn
\\
A_{4,\mu}(X)&=&-2\6_{\mu} E_1(X)-2E_{6,\mu}(X). \nn 
\end{eqnarray}
Eliminating the coefficients one finds that (\ref{two}) + (\ref{three}) can 
be expressed by
\begin{eqnarray}
\label{twothree}
&
\sigma_{0}\(\eta(\6\ep)^2 (B_5(X)-A_1(X))+\eta\ep\6^2\ep E_1(X)+
\ep\6\ep\6\eta A_1(X)-2\ep\6\ep\5\eta\5\cD x^{\mu} E_{6,\mu}(X)\) \nn
&
\\
&
+\sigma_{0,1}\(-\eta\6\eta\6\ep\psi^\mu\6_{\mu} A_1(X)-
2\5\eta\eta\6\ep\5\cD x^{\mu}\psi^\nu\6_{\nu} E_{6,\mu}-
\5\eta\eta\6\ep\5\cD x^{\rho}\psi^\nu\Omega_{\rho\nu}{}^\mu E_{6,\mu}\),
&
\end{eqnarray}
where we have used the on-shell equality (\ref{Sigma13}). Next we consider 
the terms involving derivatives of $\5\eta$
\begin{eqnarray}
\label{four}
&
\ep\6^2\ep\eta\5\6\5\eta A_2(X)+
(\6\ep)^2\eta\5\6\5\eta B_2(X)+
\ep\6\ep\5\eta\5\6^2\5\eta B_6(X)
& \nn
\\
&
+\ep\6\ep\6\eta\5\6\5\eta B_7(X)+
\ep^2\6^2\eta\5\6\5\eta E_3(X)+
\ep\6\ep\5\6\5\eta\5\eta\5\cD x^{\mu} C_{5,\mu}(X), 
&
\end{eqnarray}
which implies via (\ref{sex})
\begin{eqnarray}
&
B_7(X)=0,\quad A_2(X)=B_6(X)=B_2(X)=-2 E_3(X),
& \nn
\\
&
C_{5,\mu}(X)=-\6_\mu A_2(X).
&
\end{eqnarray}
Thus (\ref{four}) can be written as
\begin{equation}
\sigma_{0}\(\ep\6\ep\5\6\5\eta A_2(X)\)-
\sigma_{0,1}\(\6\ep\5\6\5\eta\eta\psi^\mu\6_{\mu} A_2(X)\)
\end{equation}
and thus be removed from $\omega_{2}$. In the last step we consider 
contributions containing $U(1)$ ghosts, i.e.
\begin{eqnarray}
\label{five}
&
\ep\6^2\ep\eta C^i A_{3,i}(X)+
(\6\ep)^2\eta C^i B_{3,i}(X)+
\ep\6\ep\6\eta C^i B_{8,i}(X)+
\ep^2\6^2\eta C^i E_{4,i}(X)
& \nn
\\
&
+\ep\6\ep\5\6\5\eta C^i B_{9,i}(X)+
\ep\6\ep C^i\5\eta\5\cD x^{\mu} C_{6,\mu i}(X)+
\ep\6\ep C^i C^j B_{10,ij}(X).
&  
\end{eqnarray}
(\ref{sex}) imposes $B_{10,ij}(X)=B_{9,i}(X)=B_{8,i}(X)=0$. Furthermore we 
derive the conditions
\begin{equation}
A_{3,i}(X)=B_{3,i}(X)=-2 E_{4,i}(X) \quad C_{6,\mu i}(X)=-\6_{\mu}A_{3,i}(X).
\end{equation}
Using the on-shell equality (\ref{Sigma12}), 
(\ref{five}) can be written as
\begin{eqnarray}
&\sigma_{0}\(\ep\6\ep C^i A_{3,i}(X)\)& \nn
\\
&+\sigma_{0,1}\(\6\ep C^i\eta\psi^\mu\6_{\mu}A_{3,i}(X)-
\6\ep\eta\5\eta\psi^\mu\bar\cD x^{\nu}
       (\Omega_{\mu\nu i}-\Omega_{\mu\nu}{}^\l G_{\l i}) A_{3,i}(X)\).&
\end{eqnarray}
Hence, as in the case $g=3$ one finds
that (\ref{sex}) implies $\omega_{2}=\sigma_{0}(\dots)
+\sigma_{0,1}(\dots)$ which implies (\ref{H06}) for $g=4$.

\mysection{Analysis of Bianchi identities}
\label{gaugealgebra}

In this appendix we summarize briefly the investigation of the
Bianchi identities for two-dimensional supergravity coupled to Maxwell 
theory. The starting point is the structure equation
\begin{equation}\label{neugrcomm}
[\cD_{A},\cD_{B}\}=-{T_{AB}}^C\cD_{C}
-R_{AB}\delta_L
-{F_{AB}}^i\delta_i,
\end{equation} 
where $[\cdot,\cdot\}$ denotes the graded commutator,
$\{\cD_{A}\}=\{\cD_{a},\cD_\alpha\}$
contains the covariant 
derivatives $\cD_a$ and
covariant supersymmetry transformations $\cD_\alpha$,
$\delta_L=(1/2)\ep^{ab}l_{ab}$ is the Lorentz generator
and $\delta_i$ are the $U(1)$ 
generators (represented trivially in our case). The ``torsions''
${T_{AB}}^C$, ``curvatures'' $R_{AB}$ and
``field strengths'' ${F_{AB}}^i$ are
generically field dependent and determined from the
Bianchi identities implied by (\ref{neugrcomm}).
Using the constraints (\ref{gravconstr}) and 
(\ref{u1constr}) one obtains for the 
torsions
\begin{eqnarray}
  \label{torsions}
  T_{\a\b}{}^a &=& 2\Ii(\g^aC)_{\a\b} \nn\\
  T_{a\b}{}^\a &=& \sfrac{1}{4}S(\g_a)_\b{}^\a \nn\\
  T_{ab}{}^\a &=& \sfrac{\Ii}{4}\ep_{ab}(C\g_*)^{\a\b}\cD_{\b}S,  
\end{eqnarray}
where $S$ is the auxiliary scalar field of the gravitational multiplet. For 
the curvatures one obtains
\begin{eqnarray}
  \label{curvatures}
  R_{\a\b} &=& \Ii S (\g_*C)_{\a\b} \nn\\
  R_{a\a} &=& \sfrac{\Ii}{2}(\g_a\g_*)_{\a}{}^\b\cD_{\b}S \nn\\
  R_{ab} &=& \sfrac{1}{4}\ep_{ab}(S^2+\cD^2 S),
\end{eqnarray}
and the field strengths are given by
\begin{eqnarray}
  \label{fieldstr}
  F_{\a\b}{}^i &=& 2\Ii(\g_*C)_{\a\b}\Ph^i \nn\\
  F_{a\a}{}^i &=& (\g_a)_\a{}^\b\lambda_\b^i\ . 
\end{eqnarray}
The supersymmetry transformations of $\lambda^i_\b$
and $F_{ab}^i$ turn out to be
\begin{eqnarray}
  \label{susyu1}
  \cD_\a\lambda^i_\b &=& \Ii(\g^a\g_*C)_{\a\b}\cD_a\Ph^i
                       +\sfrac{\Ii}{2}(\g_*C)_{\a\b}\ep^{ab}F_{ba}^i
                       +\sfrac{\Ii}{2}(\g_*C)_{\a\b}S\Ph^i \nn \\
  \cD_\a F_{ab}^i &=& -(\g_b\cD_a\lambda^i)_\a+(\g_a\cD_b\lambda^i)_\a
                      +\sfrac{1}{2}\ep_{ab}\cD_\a S \Ph^i
                      +\sfrac{1}{2}\ep_{ab}S(\g_*)_\a{}^\delta\lambda_\delta^i.
\end{eqnarray}
Introducing the corresponding connection 1-forms and proceeding along the 
lines of~\cite{Brandt:1997mh} one identifies the covariant 
derivatives $\cD_a$ in terms of partial derivatives and
connections, and the curvatures,
field strengths and torsions with two lower Lorentz 
indices in terms of the connections and the other field strengths. 
Owing to the constraint $T_{ab}{}^c=0$ this yields
the expression (\ref{spinconnection}) for the spin connection.
Furthermore one obtains
\[
F_{ab}{}^i = E_a{}^n E_b{}^m(\partial_n A_m^i-\partial_m A_n^i
             -(\chi_m\g_n\lambda^i)+(\chi_n\g_m\lambda^i)
             -2\Ii(\chi_m\g_*C\chi_n)\Ph^i)
\]
and the expression for $T_{ab}{}^{\a}$ can be used to express the 
supersymmetry transformation of the auxiliary field $S$ as
\[
\cD_\a S = 4\Ii(\g_*C)_{\a\b}\ep^{nm}\nabla_m\chi_n{}^\b 
           -\Ii(\g^mC)_{\a\b}\chi_m{}^\b S.
\]
The full BRST transformations~(\ref{sugra}), (\ref{matter}) and (\ref{U1}) are 
then obtained by adding the Weyl transformations by hand and imposing $s^2=0$ 
on all fields. To achieve this in an off-shell setting,
one introduces the super-Weyl symmetry on the gravitino and the gaugino 
and the local shift symmetry of the 
auxiliary field $S$.

\mysection{BRST transformations of superconformal tensor fields}
\label{transformations1}

This appendix collects the BRST transformations of the 
superconformal tensor fields and corresponding ghost variables
derived in section \ref{covfields}. The transformations of the 
undifferentiated fields read
\beann
s\eta &=&\eta\6\eta-\ep\ep
\\
s\5\eta &=&\5\eta\5\6\5\eta-\5\ep\5\ep
\\
s\ep &=& \eta\6\ep-\sfrac 12\ep\6\eta
\\
s\5\ep &=& \5\eta\5\6\5\ep-\sfrac 12\5\ep\5\6\5\eta
\\
sC^i &=& \eta\5\eta \cF^i+\eta\5\ep\lambda^i
         +\5\eta\ep\5\lambda^i+\ep\5\ep\hat\phi^i
\\
sX^M &=& (\eta\cD+\5\eta\5\cD)X^M+\ep \psi^M+\5\ep \5\psi^M
\\
s\psi^M &=& (\eta\cD+\5\eta\5\cD)\psi^M +\sfrac 12\6\eta \psi^M 
            +\ep\cD X^M-\5\ep \hat F^M
\\
s\5\psi^M &=& (\eta\cD+\5\eta\5\cD)\5\psi^M +\sfrac 12\5\6\5\eta \5\psi^M 
            +\5\ep\5\cD X^M+\ep \hat F^M
\\
s\hat F^M &=&(\eta\cD+\5\eta\5\cD)\hat F^M+\sfrac 12(\6\eta+\5\6\5\eta)\hat F^M
         +\ep\cD\5\psi^M-\5\ep\5\cD\psi^M
\\
s\hat\phi^i &=&(\eta\cD+\5\eta\5\cD)\hat\phi^i
           +\sfrac 12(\6\eta+\5\6\5\eta)\hat\phi^i
           +\ep\lambda^i+\5\ep\5\lambda^i
\\
s\lambda^i &=& (\eta\cD+\5\eta\5\cD)\lambda^i
               +(\6\eta+\sfrac 12\5\6\5\eta)\lambda^i
               +\ep\cD\hat\phi^i+\5\ep \cF^i+\6\ep\hat\phi^i
\\
s\5\lambda^i &=& (\eta\cD+\5\eta\5\cD)\5\lambda^i
               +(\sfrac 12\6\eta+\5\6\5\eta)\5\lambda^i
               +\5\ep\5\cD\hat\phi^i-\ep \cF^i+\5\6\5\ep\hat\phi^i
\\
s \cF^i &=& (\eta\cD+\5\eta\5\cD) \cF^i
          +(\6\eta+\5\6\5\eta) \cF^i
          -\ep\cD\5\lambda^i+\5\ep\5\cD\lambda^i
          -\6\ep\5\lambda^i+\5\6\5\ep\lambda^i
\eeann
The $s$-transformations of covariant $\cD$ or $\5\cD$ derivatives (of first 
or higher order) of a field are obtained by applying $\cD$'s and/or $\5\cD$'s 
to the transformations given above, 
using the rules $\cD\eta=\6\eta$, $\cD\5\eta=0$, 
$\cD\ep=\6\ep$, $\cD\5\ep=0$ etc, as well as $[\cD,\5\cD]=0$. 
E.g., one gets
\beann
s\cD X^M &=& (\eta\cD+\5\eta\5\cD)\cD X^M
             +\6\eta\cD X^M+\ep\cD\psi^M+\5\ep\cD\5\psi^M
             +\6\ep\psi^M
\\
s\5\cD X^M &=& (\eta\cD+\5\eta\5\cD)\5\cD X^M
             +\5\6\5\eta\5\cD X^M+\ep\5\cD\psi^M+\5\ep\5\cD\5\psi^M
             +\5\6\5\ep\5\psi^M
\\
s\cD\5\cD X^M &=& (\eta\cD+\5\eta\5\cD)\cD\5\cD X^M
             +(\6\eta+\5\6\5\eta)\cD\5\cD X^M
\\
& &          +\ep\cD\5\cD\psi^M+\5\ep\cD\5\cD\5\psi^M
             +\6\ep\5\cD\psi^M+\5\6\5\ep\cD\5\psi^M
\\
s\cD\psi^M &=& (\eta\cD+\5\eta\5\cD)\cD\psi^M
               +\sfrac 32\6\eta\cD\psi^M+\sfrac 12\6^2\eta\psi^M
\\
& &            +\ep\cD^2 X^M+\6\ep\cD X^M-\5\ep\cD \hat F^M
\\
s\5\cD\5\psi^M &=& (\eta\cD+\5\eta\5\cD)\5\cD\5\psi^M
               +\sfrac 32\5\6\5\eta\5\cD\5\psi^M
               +\sfrac 12\5\6^2\5\eta\5\psi^M
\\
& &            +\5\ep\5\cD^2 X^M+\5\6\5\ep\5\cD X^M+\ep\5\cD \hat F^M
\\
s\5\cD\psi^M &=& (\eta\cD+\5\eta\5\cD)\5\cD\psi^M
                 +\sfrac 12\6\eta\5\cD\psi^M+\5\6\5\eta\5\cD\psi^M
\\
& &              +\ep\cD\5\cD X^M-\5\6\5\ep \hat F^M-\5\ep\5\cD \hat F^M
\\
s\cD\5\psi^M &=& (\eta\cD+\5\eta\5\cD)\cD\5\psi^M
                 +\6\eta\cD\5\psi^M+\sfrac 12\5\6\5\eta\cD\5\psi^M
\\
& &              +\5\ep\cD\5\cD X^M+\6\ep \hat F^M+\ep\cD \hat F^M
\eeann

\mysection{BRST transformations 
of superconformal antifields}\label{transformations2}

In this appendix we present
the full $s$ transformations of the superconformal antifields
associated with the matter and gauge multiplets, 
using the following notation: 
\beann
G_{MN}&:=&H_{(MN)}(X)
\\
D_i&:=&D_i(X)
\\
\Omega_{KNM}&:=&\6_K H_{MN}(X)-\6_M H_{KN}(X)+\6_N H_{KM}(X)
\\
&=&2\Gamma_{KNM}-H_{KNM}\quad (H_{KNM}=3\6_{[K}H_{NM]})
\\
R_{KLMN}&:=&\6_M\6_{[K} H_{L]N}(X)-\6_N\6_{[K} H_{L]M}(X)
\\
&=&\sfrac 12\, (\6_K\Omega_{LMN}-\6_L\Omega_{KMN})
=\sfrac 12\, (\6_M\Omega_{KNL}-\6_N\Omega_{KML}).
\eeann
$\Omega_{KNM}$ and $R_{KLMN}$ enjoy the following properties:
\beann
&\Omega_{KMN}+\Omega_{KNM}=\Omega_{MKN}+\Omega_{NKM}=2\6_K G_{MN}&\\
&R_{KLMN}=-R_{LKMN}=-R_{KLNM}\ ,\quad \6_{[J}R_{KL]MN}=0.&
\eeann
The full BRST transformations of the undifferentiated superconformal 
matter antifields are
\beann
s F^*_M &=& -\hat\phi^i\6_M D_i+2G_{MN}\hat F^N+\psi^K\5\psi^N\Omega_{KNM}
\\
         & & +(\eta\cD+\5\eta\5\cD)F^*_M
             +\sfrac 12(\6\eta+\5\6\5\eta)F^*_M
             -\ep\5\psi^*_M+\5\ep\psi^*_M
\\
s\psi^*_M &=& \5\lambda^i\6_M D_i+\5\psi^N\hat\phi^i\6_N\6_M D_i
              +2G_{MN}\5\cD\psi^N
\\
         & & +\5\cD X^N\psi^K\Omega_{KNM}-\hat F^N\5\psi^K\Omega_{MKN}
             -\psi^K\5\psi^N\5\psi^L R_{KMLN}
\\
         & & +(\eta\cD+\5\eta\5\cD)\psi^*_M
             +(\sfrac 12\6\eta+\5\6\5\eta)\psi^*_M
             +\ep X^*_M+\5\ep\5\cD F^*_M+\5\6\5\ep F^*_M
\\
s\5\psi^*_M &=& -\lambda^i\6_M D_i-\psi^N\hat\phi^i\6_N\6_M D_i
              +2G_{MN}\cD\5\psi^N
\\
         & & +\cD X^N\5\psi^K\Omega_{NKM}+\hat F^N\psi^K\Omega_{KMN}
             +\psi^K\psi^L\5\psi^N R_{LKMN}
\\
         & & +(\eta\cD+\5\eta\5\cD)\5\psi^*_M
             +(\6\eta+\sfrac 12\5\6\5\eta)\5\psi^*_M
             +\5\ep X^*_M-\ep\cD F^*_M-\6\ep F^*_M
\\
sX^*_M   &=& -2G_{MN}\cD\5\cD X^N-\cD X^K\5\cD X^L\Omega_{KLM}
             +\hat F^K\hat F^L\Omega_{MKL}
\\
         & & +\cD\5\psi^K\5\psi^L\Omega_{MLK}
             -\psi^K\5\cD\psi^L\Omega_{KML}
\\
         & & +\cD X^N\5\psi^K\5\psi^L R_{NMLK}
             +\5\cD X^N\psi^K\psi^L R_{LKNM}
\\
         & & +\hat F^N\psi^K\5\psi^L\6_M\Omega_{KLN}
             +\sfrac 12\psi^R\psi^K\5\psi^N\5\psi^L\6_MR_{KRLN}
\\
         & & +\cF^i\6_MD_i-(\psi^N\5\lambda^i-\5\psi^N\lambda^i
             +\hat F^N\hat\phi^i+\psi^N\5\psi^K\hat\phi^i\6_K)\6_N\6_MD_i
\\
         & & +(\eta\cD+\5\eta\5\cD)X^*_M+(\6\eta+\5\6\5\eta)X^*_M
\\
         & & +\ep\cD\psi^*_M+\5\ep\5\cD\5\psi^*_M
             +\6\ep\psi^*_M+\5\6\5\ep\5\psi^*_M
\eeann
The $s$ transformation of the superconformal antifields for the gauge 
multiplet read 
\beann
s\lambda^*_i&=& \5\psi^M\6_MD_i
\\
         & & +(\eta\cD+\5\eta\5\cD)\lambda^*_i
             +\sfrac 12\5\6\5\eta\lambda^*_i
             +\ep\phi^*_i-\5\ep \5A^*_i
\\
s\5\lambda^*_i&=& -\psi^M\6_MD_i
\\
         & & +(\eta\cD+\5\eta\5\cD)\5\lambda^*_i
             +\sfrac 12\6\eta\5\lambda^*_i
             +\5\ep\phi^*_i-\ep A^*_i
\\
s\phi^*_i&=& -\hat F^M\6_M D_i-\psi^M\5\psi^N\6_N\6_M D_i
\\
         & & +(\eta\cD+\5\eta\5\cD)\phi^*_i
             +\sfrac 12(\6\eta+\5\6\5\eta)\phi^*_i
\\
         & & +\ep\cD\lambda^*_i+\5\ep\5\cD\5\lambda^*_i
             -\ep\5\ep C^*_i
\\
sA^*_i   &=& -\cD X^M\6_M D_i
\\
         & & +(\eta\cD+\5\eta\5\cD)A^*_i+\6\eta A^*_i
\\
         & & +\5\ep\cD\lambda^*_i-\ep\cD\5\lambda^*_i-\6\ep\5\lambda^*_i
             -\5\ep\5\ep C^*_i
\\
s\5A^*_i &=& \5\cD X^M\6_M D_i
\\
         & & +(\eta\cD+\5\eta\5\cD)\5A^*_i+\5\6\5\eta \5A^*_i
\\
         & & +\ep\5\cD\5\lambda^*_i-\5\ep\5\cD\lambda^*_i
             -\5\6\5\ep\lambda^*_i-\ep\ep C^*_i
\\
sC^*_i   &=& -\cD \5A^*_i-\5\cD A^*_i
             +(\eta\cD+\5\eta\5\cD)C^*_i
             +(\6\eta+\5\6\5\eta)C^*_i
\eeann
The BRST transformations of covariant derivatives of the
covariant antifields (such as $s\cD X^*_M$) 
are obtained from the above formulae
by means of the rules described in appendix \ref{transformations1}.

\end{document}